\documentclass[useAMS,usenatbib]{mn2e}
\usepackage[pdftex]{graphicx}
\usepackage{amsmath}

\usepackage{astro_journal_names}
\defcitealias{2013JCAP...06..036T}{TZE}

\title[Fast and accurate mock catalogue generation for low-mass galaxies]{
Fast and accurate mock catalogue generation for low-mass galaxies}
\author[J. Koda et al.]{Jun Koda,$^{1,2,3}$\thanks{E-mail: jun.koda@brera.inaf.it}
  Chris Blake,$^{2}$
  Florian Beutler,$^{4}$
  Eyal Kazin,$^{2,3}$ and
  Felipe Marin$^{2,3}$\\
  $^1$ INAF -- Osservatorio Astronomico di Brera, via E. Bianchi 46,
       23807 Merate, Italy\\
  $^2$ Centre for Astrophysics \& Supercomputing, Swinburne University of 
       Technology, PO Box 218, Hawthorn, VIC 3122, Australia  \\
  $^3$ ARC Centre of Excellence for All-sky Astrophysics (CAASTRO)\\
  $^4$ Lawrence Berkeley National Lab, 1 Cyclotron Rd, Berkeley, CA 94720, USA
}

\begin{document}
\label{firstpage}

\maketitle

\begin{abstract}
We present an accurate and fast framework for generating mock
catalogues including low-mass halos, based on an implementation of the
COmoving Lagrangian Acceleration (COLA) technique.  Multiple
realisations of mock catalogues are crucial for analyses of
large-scale structure, but conventional $N$-body simulations are too
computationally expensive for the production of thousands of
realisations.  We show that COLA simulations can produce accurate mock
catalogues with a moderate computation resource for low- to
intermediate- mass galaxies in $10^{12} M_\odot$ haloes, both in real
and redshift space.  COLA simulations have accurate peculiar
velocities, without systematic errors in the velocity power spectra
for $k \le 0.15 h\mathrm{Mpc}^{-1}$, and with only 3-per-cent error
for $k \le 0.2 h\mathrm{Mpc}^{-1}$.  We use COLA with 10 time steps and a
Halo Occupation Distribution to produce 600 mock galaxy catalogues of
the WiggleZ Dark Energy Survey.  Our parallelized code for efficient
generation of accurate halo catalogues is publicly available at
\texttt{github.com/junkoda/cola\_halo}.
\end{abstract}

\begin{keywords}
 cosmology: theory -- large-scale structure of Universe -- methods: numerical.
\end{keywords}

%
% Section 1: Introduction
%

\section{Introduction}
\label{sec:introduction}

Generating multiple realisations of mock galaxy catalogues is
essential for analysing large-scale structure in the Universe. It is a
necessary tool for evaluating the statistical uncertainties in the
clustering measurements, and systematic errors in theoretical
modelling and data analysis. The importance of accurate mock
catalogues is increasing as data analyses become more complicated and
sophisticated, and the large-scale-structure measurements become more
precise.

One of the targets of cosmological surveys is the Baryon Acoustic
Oscillation (BAO) feature imprinted in the galaxy clustering
\citep{2005MNRAS.362..505C, 2005ApJ...633..560E, 2011MNRAS.418.1707B,
  2012MNRAS.423.3430B, 2014MNRAS.441...24A}. It a `standard ruler'
that provides robust measurements of the expansion history of the
Universe through the cosmological distances as a function of
redshift. The data analysis procedure was recently refined by the
`reconstruction' technique \citep{2007ApJ...664..675E}, which improves
the precision by sharpening the BAO peak by
rewinding the large-scale displacements in part. This technique was
first applied to the Sloan Digital Sky Survey Data Release 7
\citep{2012MNRAS.427.2132P, 2012MNRAS.427.2168M}, and has become a
standard procedure \citep{2012MNRAS.427.3435A, 2014MNRAS.441...24A,
  2014MNRAS.441.3524K}.

Covariance matrices, e.g, $C_{ij} = \langle \xi(r_i) \xi(r_j) \rangle
- \langle \xi(r_i) \rangle$ $ \langle \xi(r_j) \rangle$, for two-point
correlation function $\xi(r)$, need to be calculated for any analyses
of large-scale structure to evaluate the best-fitting cosmological
parameters and their confidence regions. The ensemble averages for the
covariance matrix can be computed directly from many realisations of
mock galaxies. The benefit of multiple realisations of mock galaxy
catalogues to build the covariance matrix is not limited to BAO, but
the preference of using mocks over other methods is clear for BAO due
to its large length scale of 150 Mpc and non-trivial numerical process
in the reconstruction. Mock galaxy catalogues based on simulations can
properly evaluate the error caused by imperfect reconstruction due to
non-linear motions and realistic selection function. Alternative
methods like jack-knife sampling work for measurements on small
scales, but we often do not have enough quasi-independent subvolumes
assumed for jack-knife sampling on BAO scales. Log-Normal realisations
of the galaxy density field \citep{1991MNRAS.248....1C} can provide
many samples of large-scale fields, but non-linear dynamics is not
accurate; one of the sources of uncertainties we would like to
evaluate for the BAO measurement is the amount of non-linear motion
that is not completely captured by the rewinding in the reconstruction
algorithm.

Running many $N$-body simulations for multiple realisations is ideal,
but it requires a large amount of computation time on a
massively-parallel supercomputer. An insufficient number of mock
catalogues would give biased error evaluation; even with 600 mocks,
careful treatment is necessary to evaluate the inverse covariance matrix
\citep{2007A&A...464..399H, 2014MNRAS.439.2531P}.
Simulations will be harder as the
survey volume becomes larger \citep[BigBOSS][]{2009arXiv0904.0468S},
and the resolution required to resolve galaxy-hosting haloes become
higher, e.g., for emission-line galaxies in HETDEX
\citep{2004AIPC..743..224H}, Euclid \citep{2013LRR....16....6A}, or
FastSound \citep{2015arXiv150207900T} surveys, or for less-luminous
galaxies in deeper surveys, GAMA \citep{2011MNRAS.413..971D} or VIPERS
\citep{2014A&A...562A..23G}].
\citet{2013MNRAS.428.1036M} generated 600 mock galaxy catalogues,
`PThalos', for the Baryon Oscillation Spectroscopic Survey (BOSS),
using the 2nd-order perturbation theory and Friends-of-Friends halo
finder \citep{1985ApJ...292..371D}. Theoretical ideas of fast
simulations using analytical theories exist a decade ago
\citep{2002MNRAS.329..629S, 2002MNRAS.331..587M}, or even earlier
\citep[adhesion approximation,][]{1989MNRAS.236..385G}, but such
research attracted attention after the practical application to BOSS
\citep{2013MNRAS.433.2389M, 2013MNRAS.435..743D,
  2014MNRAS.437.2594W,2014MNRAS.439L..21K, 2014MNRAS.442.3256A,
  2015MNRAS.446.2621C, 2015MNRAS.450.1856A}. See
\citet{2015MNRAS.452..686C} for a comparison of these methods. Some of
these recent methods randomly generate haloes, instead of resolving
haloes, using a probability that depends on the local dark matter
density.

We generate 600 mock galaxy catalogues for the WiggleZ Dark Energy
Survey \citep{2010MNRAS.401.1429D} for the improved BAO measurement
using the reconstruction technique \citep{2014MNRAS.441.3524K} and for
other analyses \citep{2015arXiv150203710B, 2015arXiv150603900B,
  2015arXiv150603901M}. The WiggleZ samples are emission-line galaxies
in dark-matter haloes of masses approximately $10^{12} M_\odot$, which
is about an order of magnitude smaller than the haloes hosting the
BOSS CMASS galaxies. We use the COmoving Lagrangian Acceleration
(COLA) \citep{2013JCAP...06..036T} method to run many realisations of
simulation, after finding that the PTHalo method by
\citet{2013MNRAS.428.1036M} was not able to resolve $10^{12} M_\odot$
haloes (see Section~\ref{sec:haloes}). In this paper, we present the
accuracy of COLA mocks on large scales relevant to cosmological
analyses, which was not tested with the small simulation box by
\citet{2013JCAP...06..036T}, and show that COLA is accurate not only
for massive CMASS-like galaxies \citep{2015MNRAS.452..686C} but for
lower-mass galaxies. COLA is becoming a common tool when a large
number of simulations is required \citep{2015arXiv150603737H, 2015MNRAS.449..848H,
  2015A&A...576L..17L}.

This paper is organised as follows. We first review the COLA algorithm,
and describe our COLA simulations for the WiggleZ survey in
Section~\ref{sec:cola}, and compare our simulations with
conventional $N$-body simulations in
Section~\ref{sec:accuray-simulation}. We describe our mock galaxy
catalogue based on COLA in Section~\ref{sec:mock}, and compare the
mock galaxies with those based on conventional simulations in
Section~\ref{sec:mock_accuracy}. Throughout the paper, we use a flat
$\Lambda$CDM cosmology with $\Omega_m=0.273$, $\Omega_\Lambda =
0.727$, $\Omega_b=0.0456$, $h=0.705$, $\sigma_8=0.812$, and $n_s=
0.961$, which is the WMAP5 cosmology \citep{2009ApJS..180..330K} used
for the Gigaparsec WiggleZ simulation \cite{2015MNRAS.449.1454P}.

%
% Section 2: COLA simulation
%

\section{COLA simulation}
\label{sec:cola}

We use the COmoving Lagrangian Acceleration (COLA) method invented by
\citet[][TZE hereafter]{2013JCAP...06..036T} to run many realisations
of cosmological simulations with a reasonable amount of computation
time. COLA enables a reduction in the number of time steps by
combining 2nd-order Lagrangian Perturbation Theory (2LPT) and $N$-body
simulation.

\subsection{Introduction to the COLA algorithm}
A typical time-evolution method for $N$-body simulation is the leapfrog
integration:
\begin{align}
  \label{eq:leapfrog1}
  \bmath{x}_{i+1} &= \bmath{x}_i + \bmath{v}_{i+1/2} \Delta t \\
  \label{eq:leapfrog2}
  \bmath{v}_{i+1/2} &= \bmath{v}_{i-1/2} + \bmath{F}(\bmath{x}_i) \Delta t
\end{align}
where $\bmath{x}_i$ ($i=0,1,2,\dots$) is the position of a particle at
time $t_i \equiv i\Delta t$, $\bmath{v}_{i+1/2}$ is the velocity at
$t_{i+1/2} \equiv (i+1/2)$ $\Delta t$, and $F(\bmath{x})$ is the
acceleration at $\bmath{x}$, for some time step $\Delta t$. [The
  equations are solely for illustrating the difference between the
  conventional leapfrog integration and COLA; terms for the expanding
  Universe are dropped. See, e.g., \citet{1997astro.ph.10043Q} for
  the leapfrog time stepping for cosmological simulations.] The leapfrog
integration is accurate up to second order in $\Delta t$, but the
truncation error from higher orders in $\Delta t$ makes the time
evolution inaccurate for large $\Delta t$. In addition, the time step
is usually proportional to the Hubble time $H^{-1}(t)$ to integrate
accurately in cosmological simulations, which is smaller at higher
redshifts. Since we can approximate the motion well by 2LPT at high
redshifts, we can use larger time steps at high redshifts with COLA
than the conventional leapfrog integration.

COLA has two techniques that improve the accuracy of time integration
for large time steps. First, it uses the discrete time evolution only
for the non-linear terms beyond 2LPT, i.e., the residual particle
position, velocity, and acceleration from their 2LPT contributions:
\begin{align}
  \bmath{x}_\mathrm{res} &\equiv \bmath{x} - \bmath{x}_\mathrm{2LPT}(t), \\
  \bmath{v}_\mathrm{res} &\equiv \bmath{v} - \dot{\bmath{x}}_\mathrm{2LPT}(t),\\
  \bmath{F}_\mathrm{res} &\equiv \bmath{F}(\bmath{x}) - 
                                  \ddot{\bmath{x}}_\mathrm{2LPT}(t),
\end{align}
where the dots are time derivatives, and,
\begin{equation}
  \bmath{x}_\mathrm{2LPT}(t) = \bmath{q} +
    D_1(t) \bmath{\Psi}^{(1)}(\bmath{q}) + D_2(t) \bmath{\Psi}^{(2)}(\bmath{q}),
\end{equation}
is the growing-mode solution of 2LPT, mapping the initial comoving
position $\bmath{q}$ to a later position at time $t$.  The time
evolution is given by the linear growth factor $D_1(t)$, and the
second-order growth factor $D_2(t)$, which is approximately
\footnote{The public \textsc{2LPTic} code (footnote 3), originally
  designed to generate initial conditions at high redshifts, does not
  contain the factor $\Omega^{-1/143}$, which is negligible at high
  redshift. We correctly include this factor. The effect, however, is
  negligible, only a sub-per cent contribution to the second order at
  all redshifts.}  $D_2(t) = -\frac{3}{7} D_1(t)^2
\Omega(a(t))^{-1/143}$, where $\Omega(a) = \Omega_m (\Omega_m +
\Omega_\Lambda a^3)$ is the $\Omega$ matter at scale factor $a$
\citep[see,][and references therein for 2LPT]{1995A&A...296..575B,
  2002PhR...367....1B}. The first- and second-order motions are
integrated analytically with 2LPT, which does not have the truncation
error for discrete $\Delta t$.

The second component is an \textit{ansatz} that the residual
velocity decays as,
\begin{equation}
  \label{eq:ansatz-drift}
  \bmath{v}^\mathrm{res}(t) = \bmath{v}^\mathrm{res}_{i+1/2} 
   \left(\frac{a(t)}{a(t_{i+1/2})}\right)^{n_{LPT}}
\end{equation}
for $t_i \le t \le t_{i+1}$ during a drift step $\bmath{x}_i \mapsto
\bmath{x}_{i+1}$ , and,
\begin{equation}
  \label{eq:ansatz-kick}
  \bmath{v}^\mathrm{res}(t) = A_i + B_i a(t)^{n_{LPT}}
\end{equation}
for $t_{i-1/2} \le t \le t_{i+1/2}$ during a kick step,
$\bmath{v}^\mathrm{res}_{i-1/2} \mapsto
\bmath{v}^\mathrm{res}_{i+1/2}$, where $A_i$ and $B_i$ are constants,
$a$ is the scale factor, and $n_{LPT} = -2.5$ is a free parameter.
These functions replace the linear functions of $\Delta t$ in
equations (\ref{eq:leapfrog1}-\ref{eq:leapfrog2}), and suppress the
higher-order terms. (Note that the growing mode is captured by 2LPT,
and the residual term is a decaying mode --- at least in the linear
perturbation theory.) The two ansatz are empirical, and not exactly
consistent with each other; equation~(\ref{eq:ansatz-drift}) is
assuming that $A_i$ is negligible compared to the second term with
$B_i$ in equation~(\ref{eq:ansatz-kick}). In fact,
\citetalias{2013JCAP...06..036T} suggest another ansatz,
$v^\mathrm{res}(t) = v^\mathrm{res}_{i+1/2}$, as a replacement for
equation~(\ref{eq:ansatz-drift}) for simulations starting at high
redshift $z \sim 49$ with low mass resolution, which is the other
limit that $B_i a(t)^{n_{LPT}}$ is negligible compared to $A_i$. The
optimum ansatz, including the value of $n_{LPT}$, depends on the
redshift and resolution. `Experimentation is always advised with COLA'
\citepalias{2013JCAP...06..036T}.

\subsection{Basic equations}
\label{sec:basic-eq}
We briefly review the equations of motion of dark matter particles in
the expanding Universe, and then present the COLA time evolution
equations (see also the original description by
\citetalias{2013JCAP...06..036T}).  Let $\bmath{x}$ be the comoving
coordinate of an $N$-body particle, and $\bmath{v} = a^2
\dot{\bmath{x}}$ be its canonical velocity. The canonical velocity,
$\bmath{v} = m^{-1} \partial L / \partial \dot{\bmath{x}}$, follows
from the Lagrangian,
\begin{equation}
  L = \frac{1}{2} m (a \dot{\bmath{x}})^2 - m \phi(\bmath{x}, t),
\end{equation}
where $m$ is the particle mass, $a \dot{\bmath{x}}$ is the physical
peculiar velocity, and $\phi$ is the peculiar gravitational potential
that satisfies the Poisson equation in the physical coordinate
$\bmath{\nabla}_\mathrm{phys} = \bmath{\nabla}/a$:
\begin{equation}
  \left(\frac{1}{a}\nabla\right)^2 \phi(\bmath{x}, t)= 
    4\pi G \left[\rho(\bmath{x}, t) - \bar{\rho}(t)\right],
\end{equation}
for the matter density $\rho$ and mean matter density $\bar{\rho}$. This
can be written as,
\begin{equation}
  \nabla^2 \phi (\bmath{x}, t) = 
  \frac{3}{2} H_0^2 \Omega_m a^{-1}(t) \delta(\bmath{x}, t),
\end{equation}
using the density contrast $\delta \equiv \rho/\bar{\rho} - 1$, the
present critical density $\rho_\mathrm{crit, 0} \equiv 3H_0^2/(8\pi
G)$, Hubble constant $H_0$, and the present matter density $\Omega_m
\equiv \bar{\rho}/\rho_\mathrm{crit, 0}$. The Euler-Lagrange equation
gives the equations of motion,
\begin{align}
  \label{eq:motion}
  \dot{\bmath{x}} &= \bmath{v}/a(t)^2, \\
  \dot{\bmath{v}} &= m^{-1} \partial L / \partial \bmath{x}  
                   = - \bmath{\nabla} \phi(\bmath{x}, t)
                   \equiv \bmath{F}(\bmath{x}, t).
\end{align}
We discretize the time into $n_\mathrm{step}=10$ steps, uniformly in
$a$ between scale factor 0 and 1,
\begin{align}
  a(t_i)      &= a_i      \equiv i/n_\mathrm{step}, \label{eq:drift-a}\\
  a(t_{i+1/2}) &= a_{i+1/2} \equiv (i+1/2)/n_\mathrm{step}
\end{align}
and set the initial condition,
\begin{equation}
  \label{eq:initial_time}
  \bmath{x}^\mathrm{res}(t_1) = 0, \qquad \bmath{v}^\mathrm{res}(t_{1/2}) = 0,
\end{equation}
which means that the position and the velocity are exactly equal to
those of 2LPT. This is slightly different from
\citetalias{2013JCAP...06..036T}; they set the initial condition at
$a=0.1$ for both the position and the velocity, and divide the scale
factor by 10 between $0.1$ and $1$. (Our time stepping is '9 steps' in
their language.)  Even though setting initial velocity at $t_{1/2}$ is
natural for leapfrog integration, we find that this causes 2 -- 3 per
cent excess in matter power spectrum at $k \sim 0.2 \, h
\mathrm{Mpc}^{-1}$; the original \citetalias{2013JCAP...06..036T}
initial condition may be more accurate. We present the results of the
original initial condition in
Appendix~\ref{sec:appendix_initial_time}.

The ansatz for the drift step (equation~\ref{eq:ansatz-drift}) and one of the
equations of motion (equation~\ref{eq:motion}) give,
\begin{equation}
  \label{eq:cola-drift}
  \bmath{x}^\mathrm{res}(t) 
    = \bmath{x}_i^\mathrm{res} + \bmath{v}_{i+1/2}^\mathrm{res} \int_{t_i}^t
      \left( \frac{a(t')}{a_{i+1/2}}\right)^{n_{LPT}} \frac{dt'}{a(t')^2},
\end{equation}
for $t_i \le t \le t_{i+1}$. We compute the integral numerically,
which is common for all particles. The time evolution during the kick
step (equation~\ref{eq:ansatz-kick}) is,
\begin{align}
  \label{eq:cola-kick}
  \bmath{v}^\mathrm{res}(t)
     &= \bmath{v}_{i-1/2} + \frac{a(t)^{n_{LPT}} - a_{i-1/2}^{n_{LPT}}}
             {n_{LPT} \, a(t_i)^{n_{LPT}-1} \,\dot{a}(t_i)} 
             \bmath{F}^\mathrm{res}(\bmath{x}_i).
\end{align}
for $t_{i-1/2} \le t \le t_{i+1/2}$; the constants $A_i$ and $B_i$ in
equation~(\ref{eq:ansatz-kick}) are set by matching the velocity at
$t=t_{i-1/2}$,
\begin{equation}
  \bmath{v}^\mathrm{res}(t_{i-1/2}) =  \bmath{v}^\mathrm{res}_{i-1/2},
\end{equation}
and the acceleration at $t=t_i$,
\begin{equation}
  \dot{\bmath{v}}^\mathrm{res}(t_i) = 
    B_i n_{LPT} \, a(t_i)^\mathrm{n_{LPT}-1}
    \, \dot{a}(t_i) = \bmath{F}^\mathrm{res}(x_i).
\end{equation}
We use equations (\ref{eq:cola-drift}-\ref{eq:cola-kick}) to
update the $N$-body particle positions and velocities $\bmath{x}_i
\mapsto \bmath{x}_{i+1}$, $\bmath{v}_{i-1/2} \mapsto
\bmath{v}_{i+1/2}$, and also to interpolate the quantities between
timesteps for snapshots at redshifts of our interest.

\subsection{The WiggleZ COLA (WiZ-COLA) simulation}

\begin{figure*}
  \centering
  \includegraphics[width=174mm]{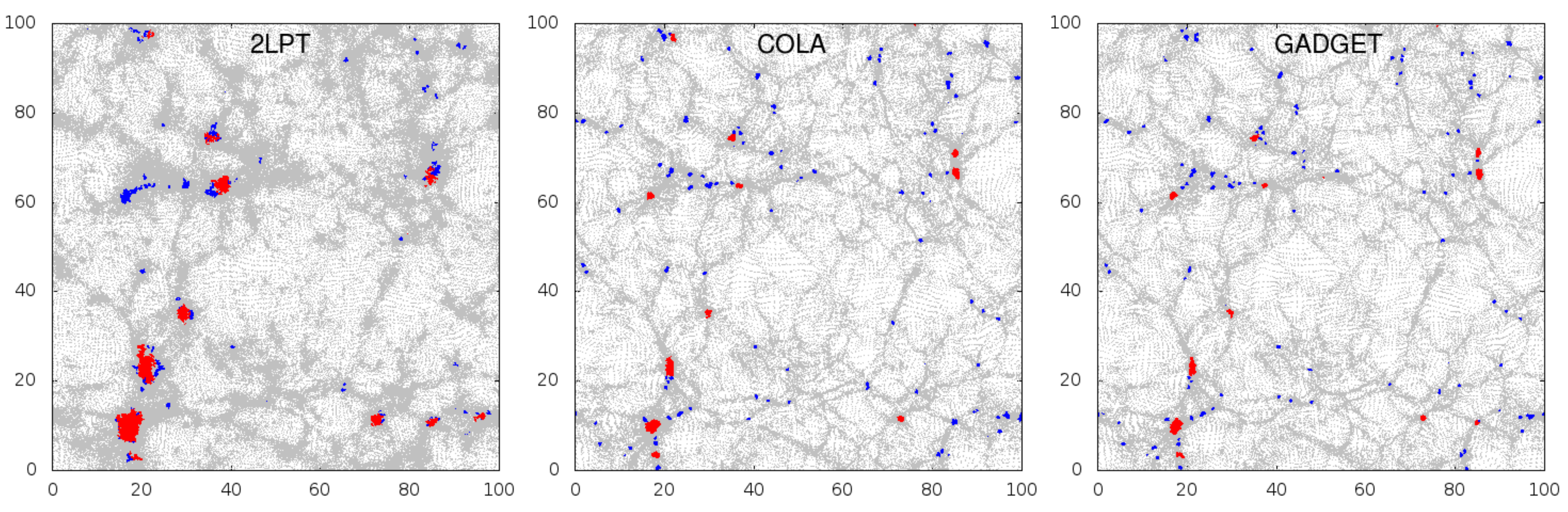}
  \caption{Simulation particles in 2LPT, COLA, and \textsc{GADGET} simulations,
    from left to right, respectively, in subvolumes of $100 \times 100
    \times 2 \, \left(h^{-1} \mathrm{Mpc} \right)^3$. The red particles
    are particles in massive haloes, $M \ge 10^{13} \,h^{-1} M_\odot$, and
    blue particles are in low-mass haloes, $10^{12} \,h^{-1} M_\odot \le M
    < 10^{13} \,h^{-1} M_\odot$. 2LPT simulation can only resolve
    massive haloes, while COLA can resolve both massive and low-mass
    haloes.}
  \label{fig:particles}
\end{figure*}

The WiggleZ-COLA (WiZ-COLA) simulation is a set of COLA simulations
designed for the WiggleZ Dark Energy Survey
\citep{2010MNRAS.401.1429D} to quantify the systematic and statistical
errors in data analyses. We run 3600 COLA simulations with different
initial random modes to generate 600 independent realisations of mock
galaxies for six survey regions in the sky (we use six independent
realisations for the six regions). The WiggleZ survey is a redshift
survey which covers about 1000 deg$^2$ up to redshift 1. The survey
volume consists of six regions in the sky, and analysed in three
redshift bins $\Delta z^\mathrm{Near}$ ($0.2 < z < 0.6$), $\Delta
z^\mathrm{Mid}$ ($0.4 < z < 0.8$), and $\Delta z^\mathrm{Far}$ ($0.6 <
z < 1.0$). We use a periodic simulation box of $600 \,
h^{-1}\mathrm{Mpc}$ on a side to cover any one of these redshift bins.
The mass of dark matter haloes hosting the emission-line galaxies in
the WiggleZ sample, inferred from the galaxy bias
\citep{2013MNRAS.432.2654M}, is about $10^{12} h^{-1} M_\odot$. We use
$1296^3$ particles, which gives the particle mass $7.5 \times 10^{9}
h^{-1}\, M_\odot$, to have more than 100 particles for haloes we need
to resolve. This mass resolution is equal to that of the Gigaparsec
WiggleZ simulation \cite[GiggleZ,][]{2015MNRAS.449.1454P}, which has
$2160^3$ particles in a $1 h^{-1}\mathrm{Gpc}$ box on a side. We use
$(3 \times 1296)^3$ meshes for Particle Mesh (PM) gravitational force
solver to resolve haloes as \citetalias{2013JCAP...06..036T}
suggested.

We parallelize the publicly available serial COLA
code\footnote{\texttt{https://bitbucket.org/tassev/colacode/}} by
\citetalias{2013JCAP...06..036T} to run simulations that satisfy the
volume and mass resolution required for the WiggleZ survey. We combine
our parallelized COLA code with a 2LPT code, \textsc{2LPTic},
\footnote{\texttt{http://cosmo.nyu.edu/roman/2LPT/}} based on
N-GenIC\footnote{\texttt{http://www.gadgetcode.org/}}, and a
Friends-of-Friends (FoF) halo finder at N-body
shop\footnote{\texttt{www-hpcc.astro.washington.edu/tools/fof.html}}
for efficient on-the-fly generation of halo catalogues. We use a
parallel Fast Fourier Transform library, FFTW3 \citep{FFTW05} for 2LPT
and PM. We follow the slab decomposition of FFTW, which slices the
volume along one axis. We divide the simulation cube into 216
equal-volume slices, and we move N-body particles between volumes
after each timestep using the Message Passing Interface (MPI). We do
not write the dark matter particles to the hard drive, we only write
the halo catalogues and the matter density field on a grid at
redshifts 0.73, 0.6, 0.44, and 0. The first three redshifts are the
effective redshifts of $\Delta z^\mathrm{Far}, \Delta z^\mathrm{Mid}$
and $\Delta z^\mathrm{Near}$, respectively.

We use 216 cores and $4 \times 216$ Gbytes of random access memory in
the Green II supercomputer at Centre for Astrophysics and
Supercomputing at Swinburne University. This number of cores is
necessary to allocate the large mesh. In Table~\ref{table:time}, we
list the composition of the computation time for one realisation; one
realisation takes about 15 minutes, and the majority of them (66 per cent)
are used for the FFTW for gravity solving. Only 2 per cent of the time
is used for 2LPT. Our COLA simulations are about a factor 50 slower
than 2LPT, but still more than 100 times faster typical $N$-body
simulations, which we describe in the following section.

\begin{table}
  \caption{Computation wallclock time of each procedure in one COLA
    simulation using 216 cores.}
  \label{table:time}
  \center
  \begin{tabular}{ccc}
    \hline
    Procedure           & Time  & Fraction\\
                        & [sec] & [per cent]\\
    \hline
    2LPT                & 18    & 2  \\
    FFT in COLA         & 583   & 66\\
    Other processes in COLA  & 114   & 13\\
    Data analysis (FOF) & 167   & 19\\
    \hline
    Total               & 882   & 100\\
    \hline
  \end{tabular}
\end{table}

%
% Section 3: Accuracy of COLA simulation
%

\section{Accuracy of COLA simulation}
\label{sec:accuray-simulation}
To test the accuracy of our COLA simulations, we compare them with
simulations performed with the same number of particles and the same
initial random modes using the publicly available Tree-PM $N$-body
code \textsc{GADGET-2} \citep{2005MNRAS.364.1105S}. For GADGET, we use
$2592^3$ PM grids and a softening length equal to 5 per cent of the
mean particle separation. We use the default values of accuracy
parameters; $\eta = 0.025$ for the time step, and $\alpha = 0.005$ for
the force accuracy. We setup the initial condition at $z=49$ using the
same 2LPT displacement fields. We make 14 realisations, and each of
the $N$-body run takes about $9000$ CPU hours using 384 computing
cores. The computation time for one realisation is about 160 times
larger than that for our COLA simulation.

\subsection{Haloes in 2LPT, COLA and GADGET simulations}
\label{sec:haloes}
In Fig.~\ref{fig:particles}, we show slices of 2LPT, COLA, and
\textsc{GADGET} simulations at redshift $0.6$. The red points are
simulation particles in `massive haloes' above $10^{13} h^{-1}
M_\odot$, and blue points are particles in `low-mass haloes' in the
range $10^{12} h^{-1} M_\odot < M < 10^{13} h^{-1} M_\odot$. We
identify the haloes with the FoF algorithm with linking length 0.2
times the mean particle separation ($\ell = 0.2$) for GADGET and COLA,
and $\ell = 0.37$ for 2LPT, following the prescription of PTHaloes by
\citet{2013MNRAS.428.1036M}. The halo masses are based on those of the
GADGET simulation. The haloes in the COLA and 2LPT simulations are
sorted by mass in descending order, and the haloes are classified to
massive or low-mass by the ranking. The massive PThaloes are found in
approximately correct locations, but low-mass PThaloes are completely
mislocated; haloes in filaments are not resolved, and the noise around
massive haloes is incorrectly identified as low mass haloes. The COLA
simulation, on the other hand, is almost indistinguishable to the
GADGET simulation; only a small number of haloes crosses the mass
boundary of $M=10^{13}h^{-1}$ due to a scatter in mass.

\begin{figure}
  \centering
  \includegraphics[width=84mm]{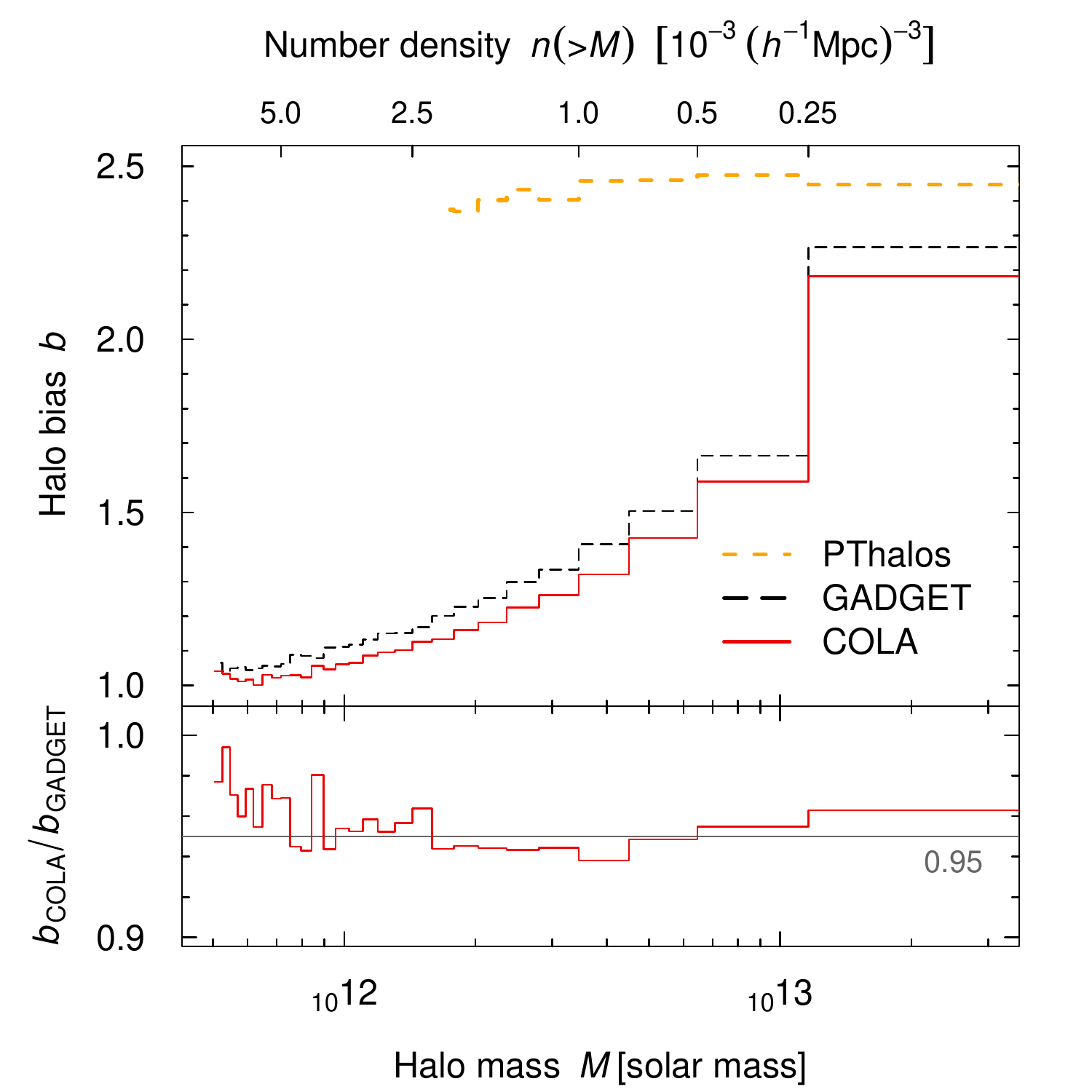}
  \caption{Linear biases of haloes grouped by their masses; each bin
    corresponds to a number density of $2.5 \times 10^{-4} (h^{-1}
    \mathrm{Mpc})^{-3}$. COLA haloes have correct bias with about 5
    per cent accuracy, while PThalos have reasonable bias only for the
    most massive bin. }
  \label{fig:halo_bias}
\end{figure}

We can also see the problem of the PTHaloes in the halo bias. In
Fig.~\ref{fig:halo_bias}, we plot the linear halo bias for haloes
grouped by their masses. Each group has a number density $2.5 \times
10^{-4} (h^{-1} \mathrm{Mpc})^{-3}$, and the corresponding mass range
is based on the GADGET simulation. The linear bias is computed by
matching the amplitude of the halo power spectrum with the matter
power spectrum of \textsc{MPTbreeze} \citep{2012MNRAS.427.2537C} for
$k \le 0.1 \,h^{-1}\mathrm{Mpc}$ (see more details in
section~\ref{sec:matter_power} for the power spectrum
computation). The bias of PTHaloes are always above 2, because all the
haloes are clustered around massive haloes. The biases of COLA haloes
have correct dependence on mass, but are 5 per cent smaller than those
of GADGET, which is probably due to the scatter in the halo
mass. Since there are more low-bias haloes than high-bias haloes,
the scatter introduces a larger fraction of low-bias haloes into the group.

\subsection{Halo mass}

\begin{figure}
  \centering
  \includegraphics[width=84mm]{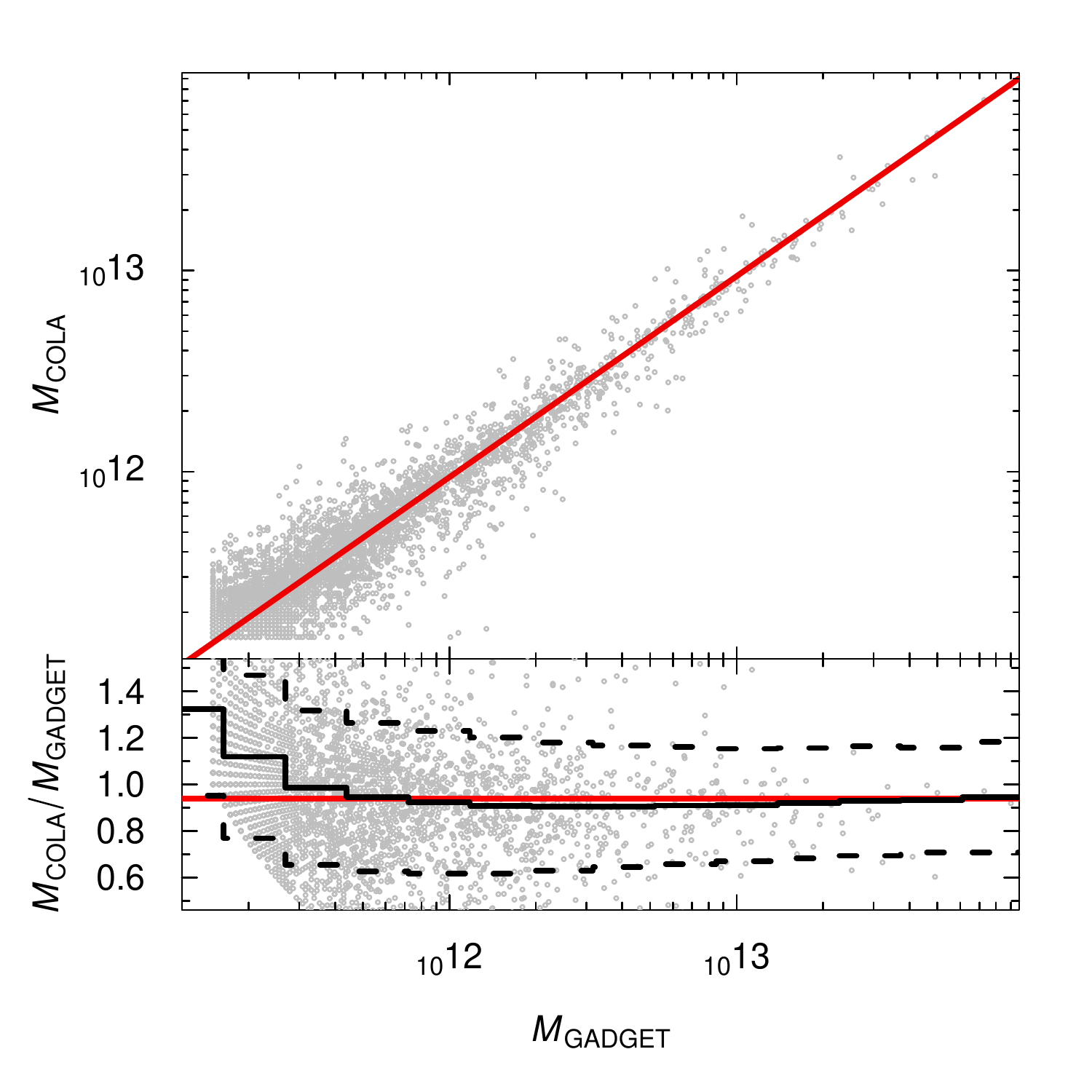}
  \caption{The relation of halo masses in the COLA simulation
    $M_\mathrm{COLA}$ and those in the GADGET simulation
    $M_\mathrm{GADGET}$. The straight red lines in both panels are the
    linear fit, $M_\mathrm{COLA} = 0.938 \, M_\mathrm{GADGET}$. The
    solid and dashed black lines in the bottom panel are the mean and
    the standard deviation of the ratio
    $M_\mathrm{COLA}/M_\mathrm{GADGET}$, respectively. The ratios are
    almost mass independent, and the scatters are 0.25 -- 0.30.}
  \label{fig:halo_matching}
\end{figure}

In Fig.~\ref{fig:halo_matching}, we plot the halo masses of COLA and
GADGET.  For each halo in the COLA simulation, $H_\mathrm{COLA}$, we
find the GADGET halo, $H_\mathrm{GADGET}$, that contains the largest
number of halo particles in $H_\mathrm{COLA}$, $f\!: H_\mathrm{COLA}
\mapsto H_\mathrm{GADGET}$, then we find the same mapping in the
opposite direction for each GADGET halo, $g\!: H_\mathrm{GADGET}
\mapsto H_\mathrm{COLA}$. In the figure, we plot the masses for a
subset of halo pairs that the both mappings exist and point to each
other:
  $\{ (H_\mathrm{COLA}, H_\mathrm{GADGET}): 
  f(H_\mathrm{COLA}) = H_\mathrm{GADGET}$ and
  $g(H_\mathrm{GADGET}) = H_\mathrm{COLA} \}$.

The linear fitting gives,
\begin{equation}
  \label{eq:mass-calibration}
  M_\mathrm{COLA} = 0.938 M_\mathrm{GADGET}.
\end{equation}
The ratio $M_\mathrm{COLA}/M_\mathrm{GADGET}$ is almost independent of
mass, except below $10^{12} h^{-1} M_\odot$ where the artificial
increase in the ratio is caused by the minimum halo mass of $32$
particles per halo. The scatters in the ratios are about 0.24 above
$10^{13} h^{-1} M_\odot$, and increase to about $0.3$ near $10^{12}
h^{-1} M_\odot$.

\subsection{Matter power spectrum}
\label{sec:matter_power}

\begin{figure}
  \centering
  \includegraphics[width=84mm]{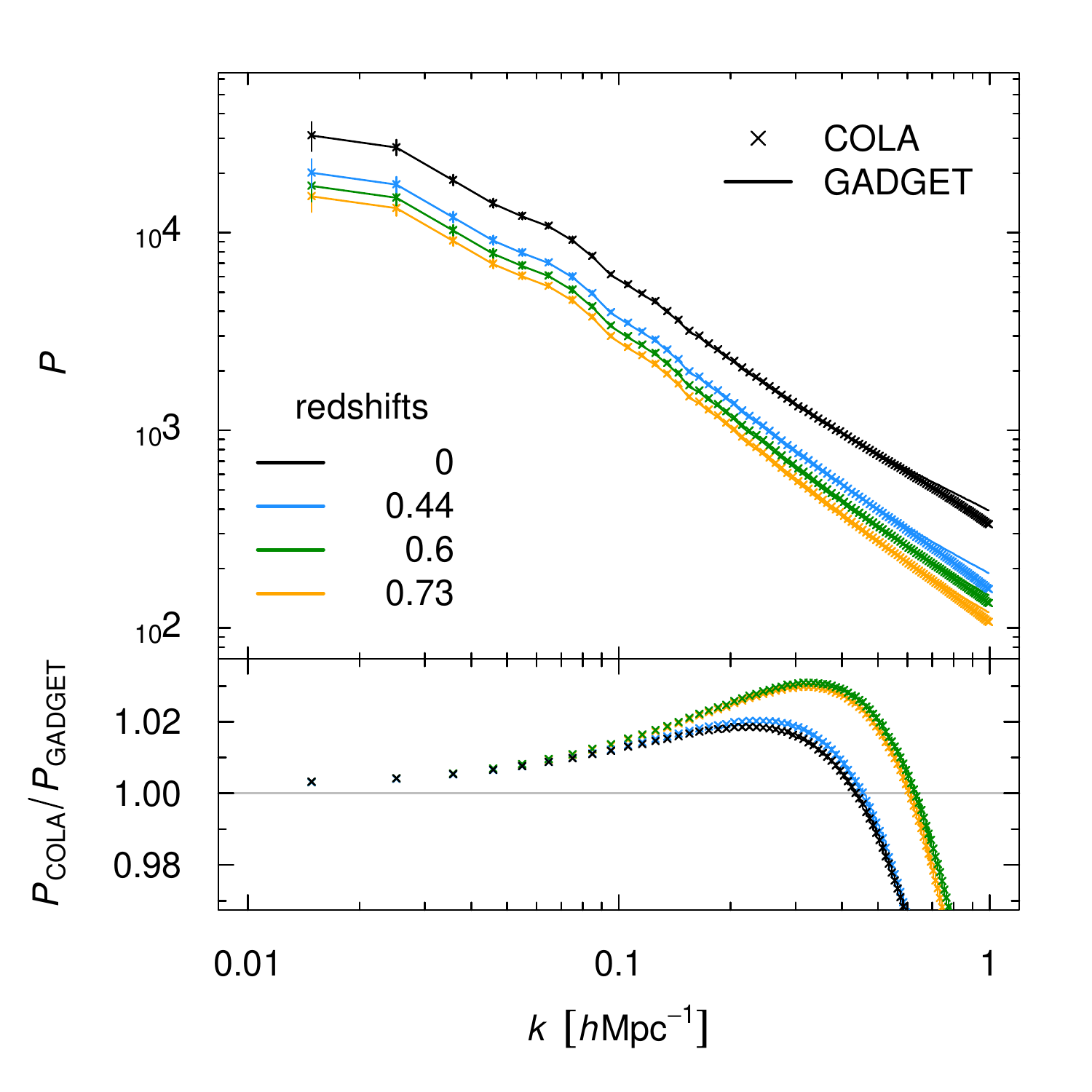}
  \caption{(\textit{Upper panel}:) Matter power spectra of a COLA
    simulation (points) and a GADGET simulation (lines) at $z=0, 0.44,
    0.6$ and $0.73$, which have the same initial
    condition. (\textit{Lower panel}:) Ratios of COLA power spectra
    to those of GADGET.}
  \label{fig:matter_power}
\end{figure}

\begin{figure}
  \centering
  \includegraphics[width=84mm]{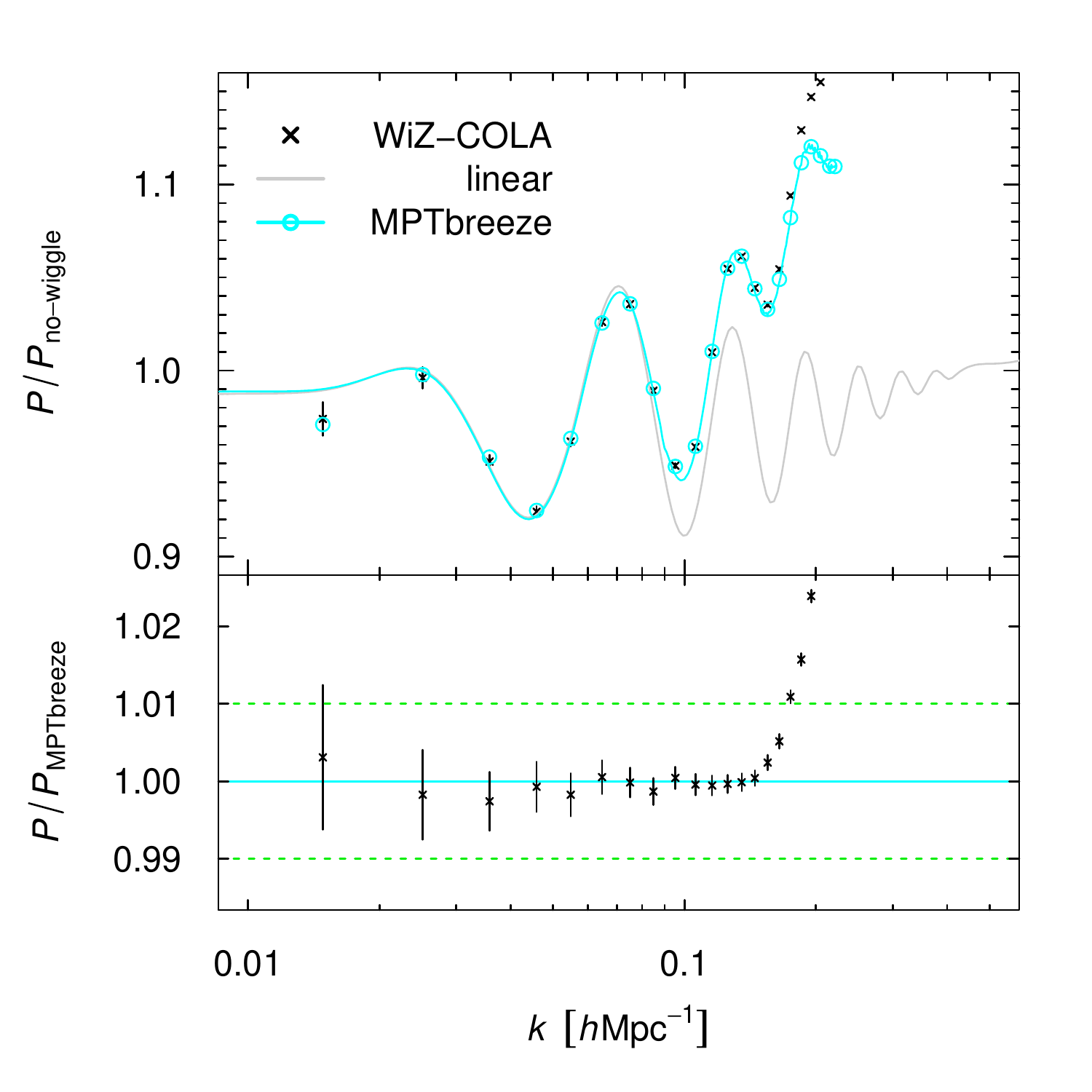}
  \caption{(\textit{Upper panel}:) Mean matter power spectrum of 3600
    COLA simulations (WiZ-COLA, black crosses) compared to a
    non-linear analytical power spectrum by MPTbreeze (cyan line) at
    $z=0.6$. Cyan circles are the analytical power spectrum averaged
    on discrete grids in Fourier space. (\textit{Lower panel}:) The
    ratio of WiZ-COLA power spectrum to the analytical power spectrum,
    both averaged on the same discrete grid in Fourier space. COLA
    simulations give very accurate overall amplitude, in agreement
    with the analytical power spectrum within statistical
    fluctuation. (The error bars are twice the standard errors in the
    mean.)}
    \label{fig:matter_power_mean}
\end{figure}

We compare the matter power spectra of COLA with those of GADGET in
Fig.~\ref{fig:matter_power} for the 14 realisations with same initial
conditions. COLA is accurate within $1.4$ per cent for $k \le 0.1
\,h \mathrm{Mpc}^{-1}$ and $2.5$ per cent for $k \le 0.2 \,h
\mathrm{Mpc}^{-1}$, respectively. The error bars are twice the
standard error in the mean,
\begin{equation}
  \label{eq:two-sigma}
  \Delta P \equiv 2 \sigma(P)/\sqrt{N_r}
\end{equation}
where $\sigma(P) = \sum_{i=1}^{N_r} (P_i - \bar{P})^2/(N_r-1)$ is the
standard deviation, $\bar{P} = \sum_{i=1}^{N_r} P_i/N_r$ is the mean,
and $N_r=14$ is the number of the realisations. The error bars for the
ratio in the bottom panel are too small to see; cosmic
variance does not directly affect the ratio of two simulations using
the same initial modes. We find an excess in the power spectrum ratio,
$P_\mathrm{COLA}/P_\mathrm{GADGET} > 1$, which was not seen in the
original paper \citepalias{2013JCAP...06..036T}; this is caused by the
slight difference in the initial condition
(equation~\ref{eq:initial_time}, see also
Appendix~\ref{sec:appendix_initial_time}). The amount of error seems
to fall into two groups; a group of redshifts 0 and 0.44, and the
other group of 0.60 and 0.73. This could be due to our interpolation
between time steps (equation~\ref{eq:cola-drift}). Redshifts 0 and
0.44 correspond to scale factor 1 and 0.694 which are close to the
drift steps (equation~\ref{eq:drift-a}), while the latter group with
slightly larger errors is away from those scale factor by about
0.025. There is probably a room for a small improvement in interpolation
formula by adding a term that uses the acceleration.

In Fig.~\ref{fig:matter_power_mean}, we plot the mean matter power
spectrum of 3600 realisations and compare with an analytical power
spectrum from \textsc{MPTbreeze} \citep{2012MNRAS.427.2537C}. The
long-wavelength modes, $ k\le 0.1 \,h\mathrm{Mpc}^{-1}$ are accurate
within the statistical uncertainty; the $\chi^2$ for the first 9 data
points, $k \le 0.1 \,h\mathrm{Mpc}^{-1}$, is 7.1. We use a publicly
available
code\footnote{\texttt{www2.yukawa.kyoto-u.ac.jp/\~atsushi.taruya/cpt\_pack.html}}
\citep{2012PhRvD..86j3528T} for the reference 'no-wiggle' power
spectrum \citep{1998ApJ...496..605E}. The good match between COLA and
MPTbreeze near $k = 0.1 \,h\mathrm{Mpc}^{-1}$ is partially due to a
coincidence, as we see errors larger than 1 per cent in
Fig.~\ref{fig:matter_power}. Here, we highlight the accuracy in the
linear growth factor in the matter power spectrum, which is a benefit
of using 2LPT in COLA; 10-time step Particle Mesh simulations, with
conventional leapfrog integration alone, have 1-2 per cent error in
the overall power spectrum amplitude \citepalias{2013JCAP...06..036T}.

The detail of calculating the power spectrum is as follows. We assign
matter densities on $324^3$ grids using the Cloud in Cell (CIC)
assignment, using all dark matter particles on the fly, and compute
the density contrast in Fourier space, $\delta(\bmath{k})$, using a
Fast Fourier Transform. The FFTW library provides discrete
$\delta(\vec{k})$ for $k_z \ge 0$ --- modes in the other half of the
Fourier space do not contain independent information due to the
reality condition $\delta(-\bmath{k}) = \delta(\bmath{k})^*$. To avoid
double counting of modes on the $k_z = 0$ plane, we use the modes
$\{k_z > 0 \} \cup \{k_z = 0 \textrm{ and } k_y > 0\} \cup \{k_z = 0
\textrm{ and } k_y = 0 \textrm{ and } k_x > 0\}$. We compute the
averages $P(k) = V^{-1} \langle \delta(\bmath{k}) \delta^*(\bmath{k})
\rangle $ and plot against the average wavenumbers $\langle k \rangle$
in bins of a fixed width $\Delta k_\mathrm{bin} = 0.01 \, h
\mathrm{Mpc}^{-1}$, where $V$ is the volume of the simulation box. The
average $\langle P \rangle$ is not an unbiased estimate of $P(\langle k
\rangle)$ in general; $P(\langle k \rangle) = \langle P(k) \rangle$ is
guaranteed only if $P(k)$ is a linear function of $k$ within the
bin. We, therefore, average the analytical power spectra on the same
discrete 3-dimensional grid for accurate comparison, which are plotted
by cyan circles in Fig.~\ref{fig:matter_power_mean}. This discrete
averaging makes statistically significant differences, especially
between $k=0.01 \,h\mathrm{Mpc}^{-1}$ and $0.02 \,
h\mathrm{Mpc}^{-1}$, where the power spectrum deviates significantly
from a linear function, reaching the maximum and turning over. We
correct for the smoothing and the aliasing effect using the procedure
by \citet{2005ApJ...620..559J}.

%
% Section 4: Halo Occupation Distribution
%

\section{Mock galaxy catalogues}
\label{sec:mock}
We populate the haloes with mock galaxies using the Halo Occupation
Distribution (HOD) prescription.

\subsection{Halo Occupation Distribution (HOD) for WiggleZ galaxies}
We use a log-normal HOD \citep{2005ApJ...630....1Z,
  2011MNRAS.412..995C} for the emission-line galaxies in the WiggleZ
sample. We assume that the probability that a dark matter halo of
mass $M$ hosts a WiggleZ galaxy is,
\begin{equation}
  P(M) = \exp \left[ -\frac{ (\log_{10} M - \log_{10} M_0)^2 }{ 2
      \sigma_{\log M}^2 } \right],
\end{equation}
where $\log_{10} M_0$ and $\sigma_{\log M}$ are parameters fitted against
data. We populate at most one galaxy per halo, without any satellite
galaxies, and set the position and velocity of the galaxy equal to the
averages of the host halo particles (i.e., the centre-of-mass position
and velocity). We do not use the error function HOD
\citep{2005ApJ...633..791Z}, because we do not expect to find
emission-line galaxies, which are young star-forming galaxies, in
groups or clusters hosted by massive haloes.

We find the two HOD parameters by matching the projected correlation
function,
\begin{equation}
  w_p(r_p) = \int_{-\pi_\mathrm{max}}^{\pi_\mathrm{max}} \xi(r_p, \pi) d\pi,
\end{equation}
with $\pi_\mathrm{max} = 60 \, h^{-1} \mathrm{Mpc}$.  We perform the
matching by populating a series of mock catalogues using a trial set
of HOD parameters, computing the mock mean, and comparing the mock mean
with the data by minimizing a $\chi^2$ statistic using a covariance
matrix obtained from jack-knife re-sampling.  Since $\log M_0$ and
$\sigma_{\log M}$ are degenerate, we fix $\sigma_{\log M} = 0.1$. We
find $\log_{10} M_0 = 12.17$ for $\Delta z_\mathrm{Near}$ and $\Delta
z_\mathrm{Far}$, and $12.28$ for $\Delta z_\mathrm{Mid}$ for FoF halo
mass $M$ without any corrections (all masses are in units of $h^{-1}
M_\odot$). COLA halo mass is about 7 per cent smaller than true
$N$-body simulation mass, but any constant calibration factor for the
mass only rescales the parameters without any change in the HOD mock.

We subsample the HOD galaxies by a realisation-independent factor to
match the smooth number density without clustering
$\bar{n}(\bmath{x})$ using the survey selection function
\citep{2010MNRAS.406..803B}. At low redshift, there are rare cases
that the number of HOD galaxies is not sufficient. In such cases, we
increase the width of the HOD $\sigma_{\log M}$ for $M < M_0$ to match
the number density, keeping the HOD same for $M > M_0$.

\subsection{HOD for BOSS CMASS galaxies}
We also generate mock catalogues for the BOSS CMASS galaxies in the
BOSS-WiggleZ overlap volume using the same simulation for the
multi-tracer analyses \citep{2015arXiv150603900B,
  2015arXiv150603901M}. We refer the reader to these papers for the
detail of the overlap regions.

We use the error function for the central galaxies, and a power law for
the satellite galaxies. We populate at most one central galaxy per
halo with a probability,
\begin{equation}
  P(M) = \frac{1}{2} \left[ 1 + \mathrm{erf}
    \left(\frac{ \log_{10} M_{200,m} - \log_{10} M_\mathrm{min} }{ \sigma_{\log M} } \right)
    \right],
\end{equation}
where $M_\mathrm{200,m}$ is a halo mass defined by the mass within a
sphere of radius $r_{200,m}$ whose mean overdensity is 200 times the
mean matter density. We denote the similar quantities for 200 times
the critical density with $M_\mathrm{200,c}$ and $r_{200,c}$.  If the
halo has a central galaxy, we draw a number of satellite galaxies from
a Poisson distribution with mean,
\begin{equation}
  \langle M_{\mathrm{sat}} \rangle = (M_{200,m}/M_0)^\beta.
\end{equation}
A satellite HOD with an additional parameter, $\left[ (M - M_1)/M_0
  \right]^\alpha$ \citep{2005ApJ...633..791Z}, is also used
frequently, but $M_1$ is usually not sensitive to the clustering data,
and does not significantly improve the fit \citep{2008MNRAS.385.1257B}.

We add a random offset and a random virial velocity to the satellite
galaxy assuming a spherical \citet*{1997ApJ...490..493N} profile,
\begin{equation}
  \rho(r) = \frac{\rho_0}{(r/r_s)(1+r/r_s)^2}.
\end{equation}
We can uniquely determine the 2-parameter profile by specifying the mass
$M_{200,c}$ and a concentration parameter $c_{200,c} =
r_{200,c}/r_s$. We draw a random concentration parameter from a known
relation in the literature, but there are several trivial steps to
convert the halo mass to an appropriate one:

\begin{enumerate}
\item We first set the FoF halo mass $M_\mathrm{FOF} = 1.066 \,
  M_\mathrm{COLA}$, which is based on our calibration between COLA and
  GADGET simulations (Fig.~\ref{fig:halo_matching});
\item compute the typical concentration factor
  $\bar{c}$ for mass $M_\mathrm{FOF}$ using \citet{2012MNRAS.423.3018P}, but
  the relation is given as a function of $M_{200,c}$;
\item convert the FoF mass halo to $M_{200,c}$ using
  \citet{2009ApJ...692..217L}, which depends on FoF mass and the
  concentration parameter. Their formula also correct for the
  resolution effect for small number of halo particles: $N_{200,c}
  \equiv M_{200,c}/m$, where $m$ is the particle mass;
\item start from an initial guess of $M_{200,c}^{(0)} = M_\mathrm{FoF}$,
  and solve steps (ii) and (iii) iteratively for mean concentration
  $\bar{c}$,
  \begin{align}
    \bar{c}_{200,c}^{(i+1)} &= \bar{c}_{200,c}(M_{200,c}^{(i)}),\\
    M_{200,c}^{(i+1)} &= M_{200,c}(M_\mathrm{FoF}, N_{200,c}^{(i)}, \bar{c}_{200,c}^{(i+1)}),
  \end{align}
  which converge quickly within several iterations;
\item draw a random concentration parameter, $\log_{10} c_{200,c}$ from
  a Gaussian distribution of mean $\log_{10} \bar{c}_{200,c}$ and standard
  deviation $\sigma_{\log c} = 0.078$
  \citep{2013MNRAS.428.1036M};
\item recompute the mass $M_{200,c}$ using the generated
  $c_{200,c}$. This determines the halo profile completely, and we can compute
  $M_{200,m}$ from the profile;
\item draw the number of central and satellite galaxies for given HOD
  parameter using $M_{200,m}$;
\item draw satellite positions from the static, spherical symmetric
  NFW profile from the phase-space distribution function. The
  static distribution function is uniquely determined from the density
  profile, assuming spherical symmetry and isotropic velocity
  distribution \citep{2004ApJ...601...37K}.
\end{enumerate}

We generate mocks for a grid of parameters, and find that $\log_{10}
M_\mathrm{min}= 12.92$, $\sigma_{\log M} = 0.31$, $\log_{10} M_0= 14.07$,
and $\beta= 1.60$, fit the projected correlation function well. Since
the HOD model contains several free parameters to fit the data, our
procedure of converting the halo mass is probably unnecessary. We also
tried a concentration parameter relation by
\citet{2001MNRAS.321..559B} with no additional scatter, but this made
little difference.

\begin{figure}
  \centering
  \includegraphics[width=84mm]{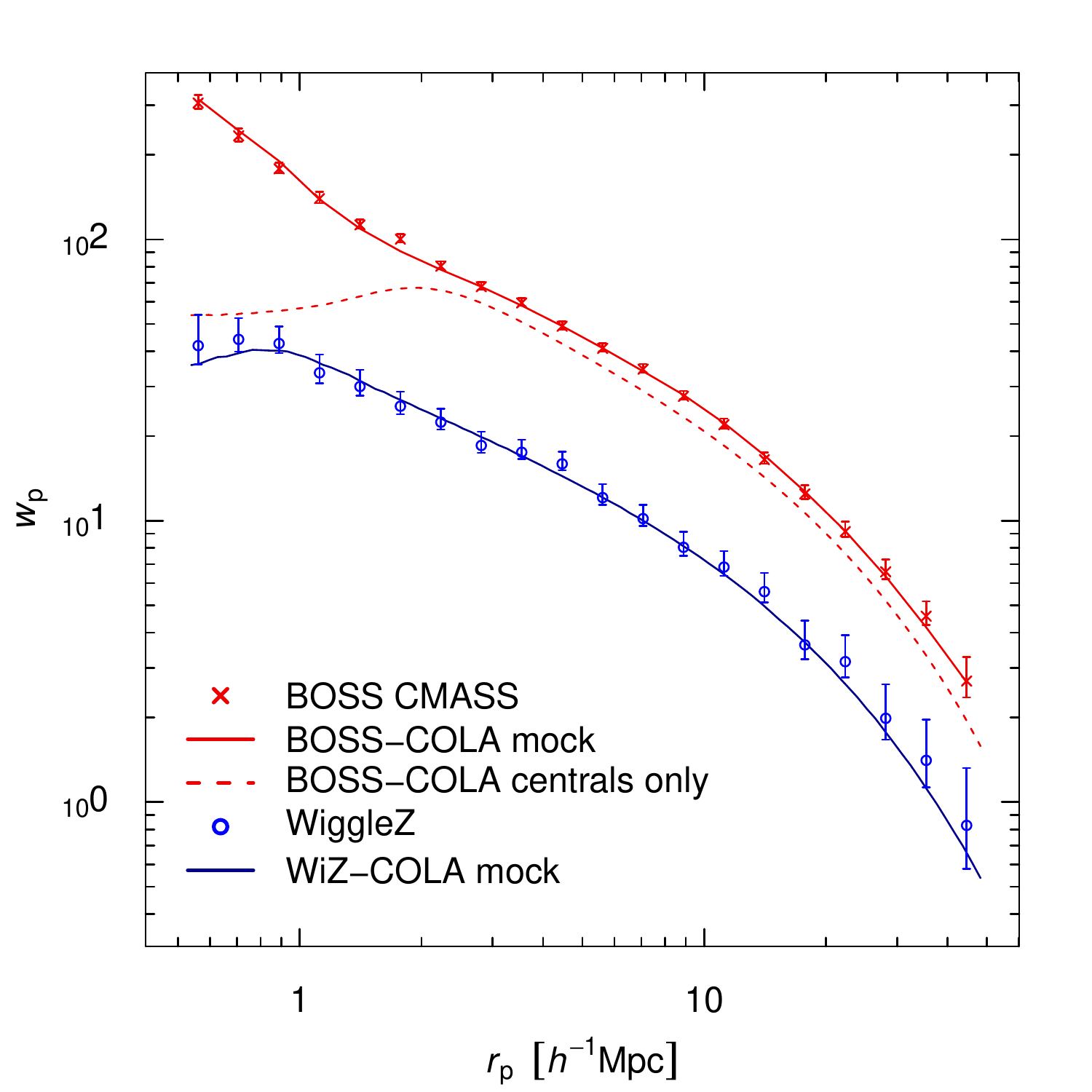}
  \caption{We tune the HOD parameters to match the projected
    correlation functions $w_p$. The mock galaxy agree with the data
    within the uncertainties. (The error bars for the data are
    $1\sigma$.)}
  \label{fig:corr_projected}
\end{figure}

In Fig.~\ref{fig:corr_projected}, we plot the projected correlation
functions for the mock and the data. The solid lines are the mean of
3600 realisations generated in the periodic box. The log-normal HOD
without satellite galaxies fits the WiggleZ data well, while a small
contribution from satellites may improve the fit for $r \simeq 0.7
h^{-1}\mathrm{Mpc}$. The BOSS CMASS galaxies clearly require
satellite galaxies for $r \simeq 2 h^{-1} \mathrm{Mpc}$.
  
\subsection{Box remapping}

\begin{table*}
  \caption{Two box configurations that we use to remap the cubic simulation
    box to cuboids, which are characterised by three integer vectors,
    $u_i$ \citep{2010ApJS..190..311C}. $L_i$ are the lengths of three
    sides of the cuboid after remapping. \label{table:remap}}
  \begin{tabular}{lcccccc}
    \hline
    Name & $\bmath{u}_1$ & $\bmath{u}_2$ & $\bmath{u}_3$
         & $L_1$ & $L_2$ & $L_3$\\
     & & & &
   [$h^{-1} \mathrm{Mpc}$] & [$h^{-1} \mathrm{Mpc}$] & [$h^{-1} \mathrm{Mpc}$]\\
    \hline
    $\sqrt{2}$ & $(1, 1, 0)$ & $(1, 0, 1)$ & $(1, 0, 0)$
               & 848.5 & 734.8 & 346.4\\
    $\sqrt{3}$ & $(1, 1, 1)$ & $(1, 0, 0)$ & $(0, 1, 0)$
               & 1039.2 & 489.9 & 424.3\\
    \hline
  \end{tabular}
\end{table*}

\begin{figure*}
  \centering
  \includegraphics[width=174mm]{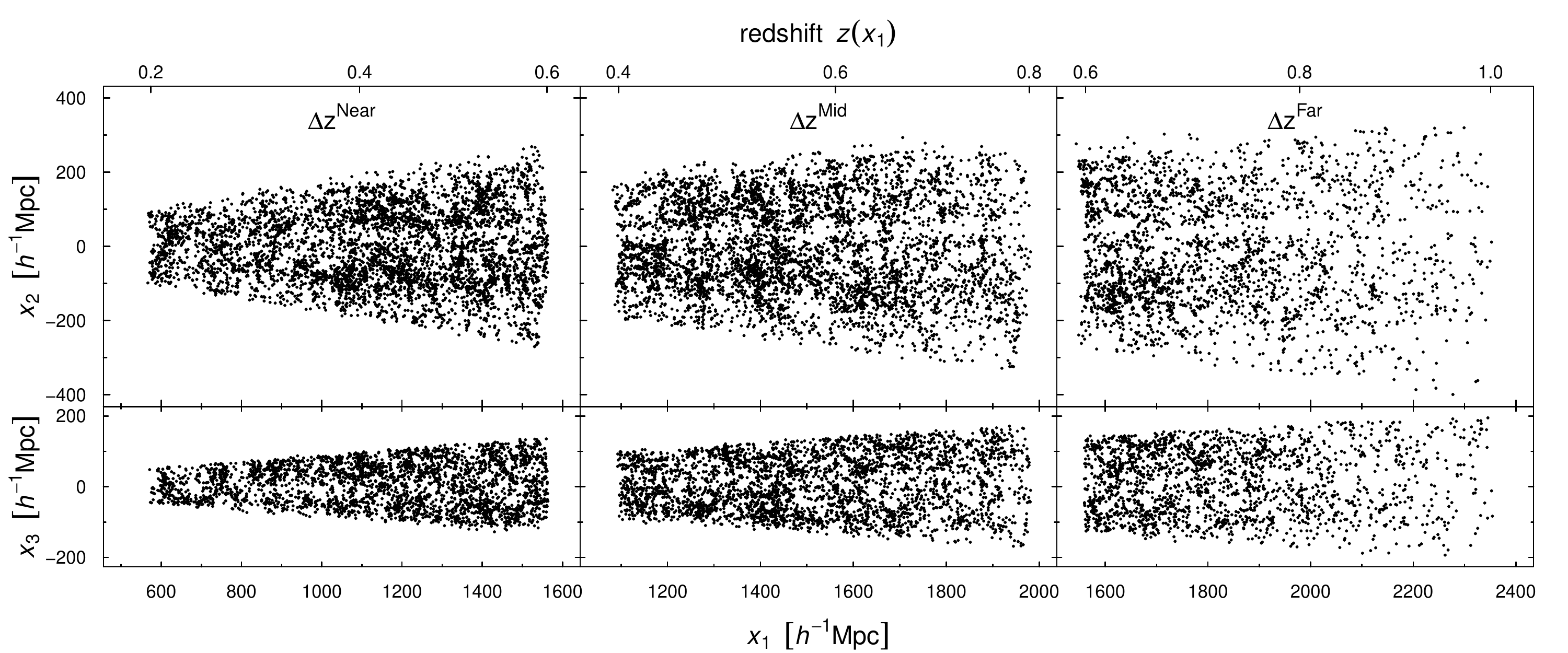}
  \caption{One realisation of the mock galaxy catalogues for the 15hr
    region. The depth of the slices is $50 \,h^{-1} \mathrm{Mpc}$.
    The coordinates are those of the remapped system, $x_i =
    \bmath{x}\cdot \bmath{e}_i$, whose origin $\bmath{x}=0$ is the
    observer.}
  \label{fig:mock}
\end{figure*}

We analyse the galaxy sample in three redshift bins, but the length
along the line of sight is still larger than the box size. We rotate
the simulation box to fit the volume with minimum overlap, using the
box remapping technique \citep{2010ApJS..190..311C} as a guide. Their
publicly available
code\footnote{\texttt{http://mwhite.berkeley.edu/BoxRemap/}} provides
a list of possible remappings from a periodic cube to cuboids. We use
two configurations, which we call $\sqrt{2}$ and $\sqrt{3}$, depending
on the size of the volume (Table~\ref{table:remap}). The
lengths of the remapped cuboid along the line of sight are, $L_1 =
\sqrt{2} L = 849 \, h^{-1} \mathrm{Mpc}$, and $L_1 = \sqrt{3} L = 1039
\, h^{-1}\mathrm{Mpc}$, respectively, where $L=600
h^{-1}\mathrm{Mpc}$ is the length of our simulation box on a side. In
the table, we list the size of the cuboid after remapping, and the
integer vectors $\bmath{u}_i$, which characterise the remapping. The
integer vectors specify the orthonormal basis of the remapped
coordinate, $\bmath{e}_i$, as follows:
\begin{align}
  \label{eq:remapping}
  \bmath{e}_1 &= \bmath{u}_1/|\bmath{u}_1| \nonumber\\
  \bmath{e}_2 &= \bmath{u}'_2/|\bmath{u}'_2|,
    \quad u'_2  \equiv \bmath{u}_2 -
    (\bmath{u}_1 \cdot \bmath{u}_2/|\bmath{u}_1|^2) \bmath{u}_1,   \nonumber \\
  \bmath{e}_3 &= \bmath{e}_1 \times \bmath{e}_2.
\end{align}
The basis vector $\bmath{e}_1$ points the line-of-sight, $\bmath{e}_2$
points the declination, and $\bmath{e}_3$ points the right ascension
directions, respectively, at the centres of the six survey regions. We
use the cuboid $\sqrt{3}$ for $\Delta z^\mathrm{Near}$, which has
enough length along the line of sight to fit the redshift range $0.2$
to $0.6$, and use $\sqrt{2}$ for $\Delta z^\mathrm{Mid}$ and $\Delta
z^\mathrm{Far}$ when we need a wider cuboid in transverse directions. A
small fraction of the survey volume was larger than the remapped
cuboid, and the same volume in the simulation box was used twice. The
fraction of such volume is 1.7 per cent of the total volume. In
Table~\ref{table:hrs}, we list the remapping we use and the fraction
of overlap for each region.

\subsection{Mock catalogue}
The overall procedure for creating a mock catalogue from a halo
catalogue is as follows:

\begin{enumerate}
 \item We fill the space with periodic replications of the simulation
   box, and rotate the positions and velocities to the remapped
   coordinate using the orthonormal basis
   (equation~\ref{eq:remapping});
 \item apply the redshift space distortion to the halo position:
  \begin{equation}
    \bmath{s} = \bmath{x} + \
                 \frac{\bmath{v} \cdot \hat{\bmath{x}}}{aH} \hat{\bmath{x}},
  \end{equation}
  where $H$ is the Hubble parameter at scale factor $a$, and
  $\hat{\bmath{x}} = \bmath{x}/|x|$ is the unit vector parallel to
  $\bmath{x}$;
 \item populate the haloes with mock galaxies using the HOD (which
   may depend on the redshift-space position at low redshift to match the
   high number density);
 \item subsample the mock galaxies to match the selection function
   (mask) of the survey. The subsample fraction is calculated to match
   the observed number of galaxies as a mean. The numbers of mock
   galaxies fluctuate around the observed number.
\end{enumerate}
For the BOSS mock, we first generate the HOD galaxies and then apply
the redshift-space distortions including the satellite virial
velocities. We can interchange the step (ii) and (iii) because we use
a position independent HOD parameters for the BOSS galaxies.  In
Fig.~\ref{fig:mock}, we plot slices of our WiZ-COLA mock catalogues
for the 15hr region.

%
% Accuracy of the HOD galaxies
%
\section{Accuracy of HOD galaxies}
\label{sec:mock_accuracy}

\begin{figure*}
  \centering
  \includegraphics[width=174mm]{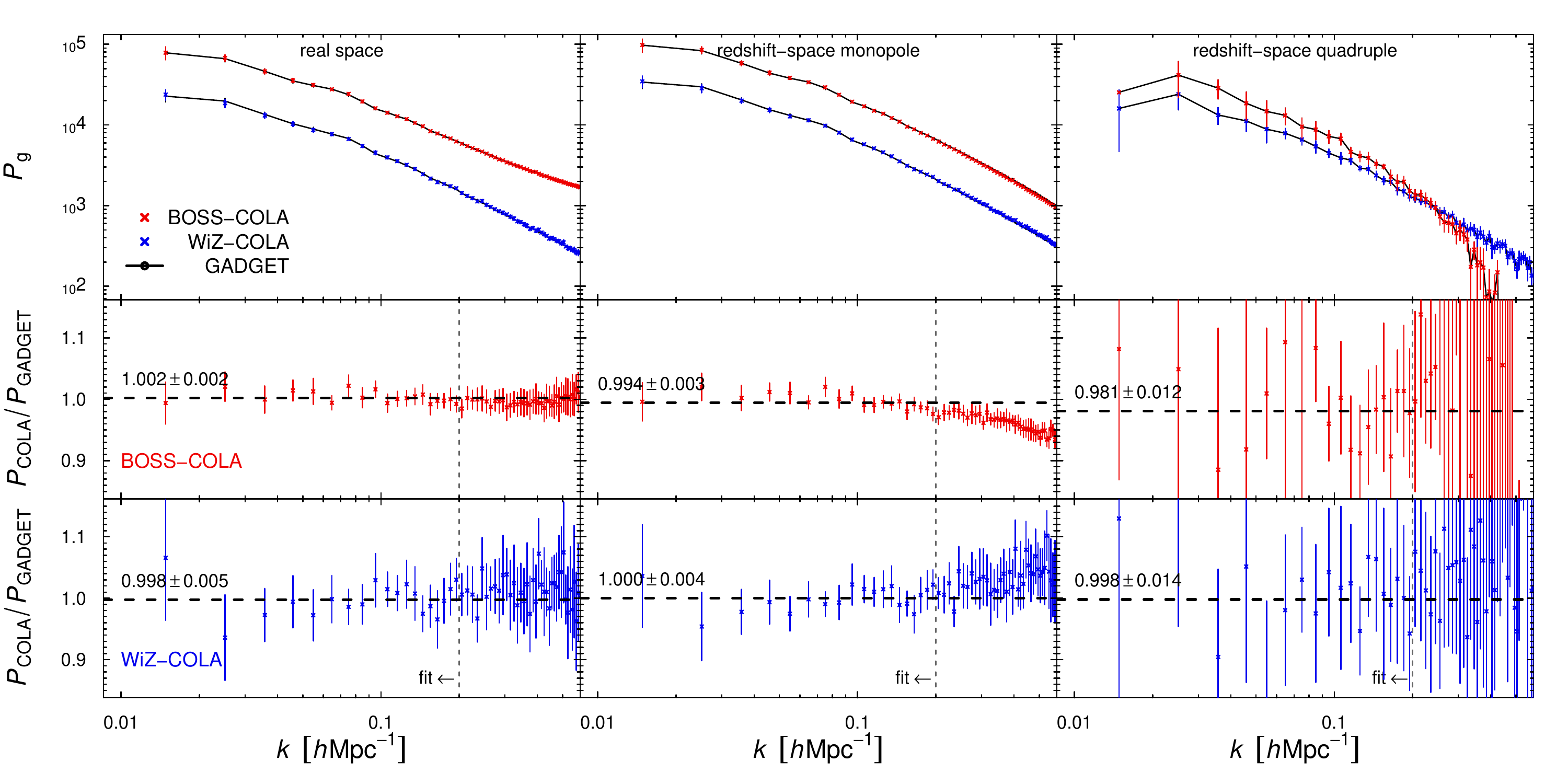}
  \caption{HOD galaxy power spectra generated from COLA versus GADGET
    in real and redshift space. COLA HOD galaxies show good
    agreement with the GADGET HOD galaxies. The horizontal lines in
    the power spectra ratios are the results of minimum $\chi^2$
    fitting, based on the diagonal errors in the ratio from 14
    realisations. The uncertainties in the fitting are 95-per cent
    intervals.}
  \label{fig:galaxy_power}
\end{figure*}

\begin{figure}
  \centering
  \includegraphics[width=84mm]{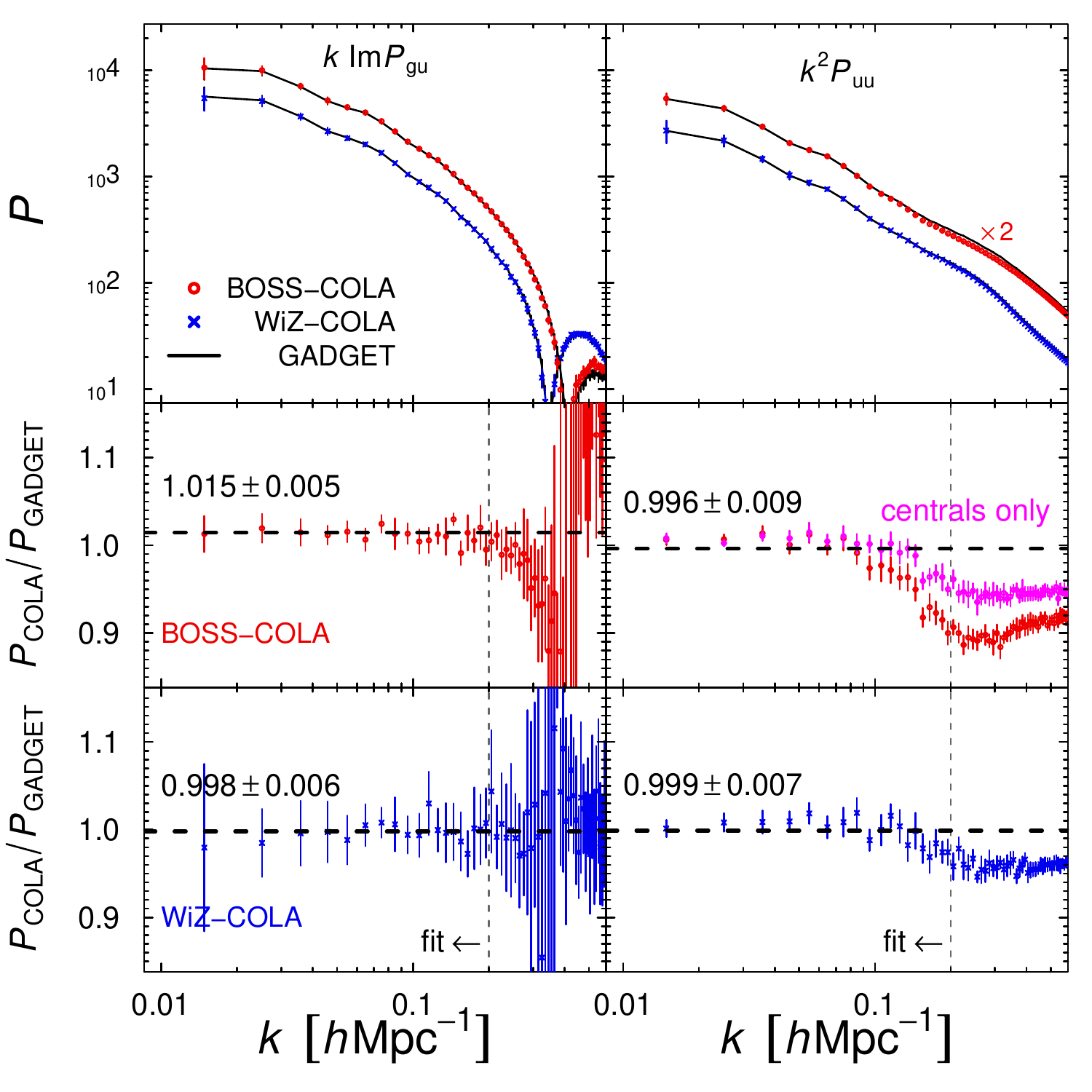}
  \caption{The cross- and auto- power spectra of HOD galaxy density
    and line-of-sight peculiar velocity. COLA has accurate peculiar
    velocities. We do not find systematic error in the velocity-galaxy
    cross power for $k\le 0.2 \,h\mathrm{Mpc}^{-1}$ and in the velocity
    auto-power for $k \le 0.15 \,h\mathrm{Mpc}^{-1}$; there are errors
    of about 3--5 per cent in the range $0.15 \,h\mathrm{Mpc}^{-1} \le k \le
    0.5 \,h\mathrm{Mpc}^{-1}$.}
  \label{fig:velocity_power}
\end{figure}

We test the accuracy of our mocks by comparing the HOD galaxies
generated from COLA with the HOD galaxies generated from GADGET N-body
simulations. We generate HOD galaxies in the periodic simulation box
and compute the power spectra. We use the HOD parameters described in
the previous section for the COLA HOD galaxies, but we determine
different HOD parameters for the GADGET haloes to match the COLA power
spectra in real space, because HOD parameters are free fitting
parameters that are usually adjusted for the observed galaxies. If we
used the same HOD parameters and the halo mass relation
(equation~\ref{eq:mass-calibration}), we would get about 5 per cent
higher galaxy power spectrum from GADGET haloes as we see in
Section~\ref{sec:accuray-simulation}, but this is not the HOD
parameters we would use. We find $\log_{10} M_0 = 12.275$ for the
WiggleZ log-normal HOD (width $\sigma_{\log M}$ is fixed to 0.1), and
$\log_{10} M_\mathrm{min} = 12.92$, $\sigma_{\log M} = 0.37$,
$\log_{10} M_0 = 14.00$, and $\beta=1.45$ for the BOSS HOD.

In Fig.~\ref{fig:galaxy_power}, we plot the power spectra in real and
redshift space. We compute the monopole ($\ell = 0$) and the
quadrupole ($\ell = 2$) moments for the redshift-space power spectrum $P^s$,
\begin{equation}
  P^s_\ell(k) = (2\ell + 1) \int P_\ell(\mu) P^s(k, \mu) d\mu,
\end{equation}
where $P_\ell$ is the Legendre polynomial, and $\mu = \hat{\bmath{k}}
\cdot \bmath{e}_3$ is the consine of the angle between the wave vector
and the fixed direction of the redshift-space distortion,
$\bmath{e}_3$, which is set to the direction of the third axis. The
procedure of computing the power spectra is the same as that in
Section~\ref{sec:haloes}; the only difference is that we also subtract
the shot noise \citep{2005ApJ...620..559J}. 

In the lower panels, we plot the ratio of the power spectra. Although
the HOD galaxies are based on simulations with the same initial
condition, the ratio of the power spectra is affected by the
randomness in populating the haloes with galaxies. The error bars are
$2\sigma$ of the mean (equation~\ref{eq:two-sigma}) based on 14
realisations. The real-space power and the redshift-space monopole are
very accurate; the ratios are consistent with unity for $k \le 0.2 \,h
\mathrm{Mpc}^{-1}$ within the statistical fluctuation, and the
statistical error is about 1 per cent. 

Since we do not have enough statistics for the quadrupole moment for
precise comparison, we also compute the cross-power spectra, $P_{gu}$,
and auto- power spectra, $P_{uu}$, between the galaxy density and the
line-of-sight peculiar velocity $u \equiv v_3$, to show the accuracy
of the peculiar velocities. The redshift-space distortion is an effect
of peculiar velocity, and the power spectrum in redshift space, $P^s$,
is approximately related to the galaxy density and velocity power
spectra in real space \citep{2004PhRvD..70h3007S},
\begin{equation}
  P^s(k, \mu) \approx P_{gg}(k) + 2 k\mu \mathrm{Im} P_{gu}(k, \mu) +
  (k\mu)^2 P_{uu}(k, \mu).
\end{equation}
(In the linear limit, the power spectra are proportional to the matter
power spectrum $P_m$, via $\mathrm{Im} P_{gu} = fb\mu P_m / k$, and
$P_{uu} = f^2 \mu^2 P_m/k^2$, respectively, where $b$ is the linear
galaxy bias and $f \equiv d \ln D_1/ d\ln a$ is the linear growth
rate.) In Fig.~\ref{fig:velocity_power}, we plot the angle-averaged
cross- and auto- power spectra, $\int_0^1 P_{gu}(k, \mu) d\mu$ and
$\int_0^1 P_{uu}(k, \mu) d\mu$.  We refer the reader to our previous
paper for technical details \citep{2014MNRAS.445.4267K}.  The cross
power spectra are also accurate with about 1 per cent scatter, but the
velocity-velocity power spectra for haloes (BOSS central galaxies and
WiggleZ galaxies) have about 3 per cent error for $k \sim 0.1 h
\mathrm{Mpc}^{-1}$, and 5 per cent error for $k \ge 0.2
h\mathrm{Mpc}^{-1}$. The BOSS satellite galaxies add additional error
due to different virial velocities caused by different HOD parameters;
this discrepancy of about 10 per cent shows that the velocity power
spectrum is sensitive to HOD parameters, in general, through the
non-linear random velocities, and is not necessarily a failure of the
COLA mocks.

A good agreement in the real-space power spectrum is not difficult to
achieve by tuning the HOD parameters or non-linear biasing models for
haloes, but such tuning does not usually work simultaneously in
redshift space. Faster mock generation techniques that uses 2LPT
usually have about 5 per cent error in the monopole and 10 per cent
error in the quadrupole of the redshift-space power spectrum
\citep{2015MNRAS.452..686C}. The primary advantage of COLA over 2LPT
based methods is the accuracy in the non-linear peculiar velocity,
which may be important for the error evaluation of BAO reconstruction,
and measurement of the growth rate. The accurate peculiar velocity is
limited to that for haloes, and we do not expect accurate densities or
virial velocities inside haloes. We find discrepancies of 10 per cent
at $k=0.1 \,h\mathrm{Mpc}^{-1}$, and 20 per cent at $k=0.2 \, h
\mathrm{Mpc}^{-1}$, respectively, in redshift-space power spectra for
$N$-body particles between COLA and GADGET, which seem to be
consequences of inaccurate virial velocities inside the haloes.

Ideally we would like to compare the accuracy of the covariance
matrix, since the main purpose of generating multiple realisations of
mock catalogues is to compute covariance, but we do not have enough
GADGET $N$-body simulations for covariance matrices. We do not have
enough realisations to compare the two-point correlation function
precisely, either. We leave these comparisons for future studies.

%% \begin{table*}
%%   \caption{$n_g$ is the HOD galaxy number density in $10^{-4} (h^{-1}
%%     \mathrm{Mpc})^{-3}$.}
%%   \label{table:z}
%%   \center
%%   \begin{tabular}{lccccc}
%%     \hline
%%        & $z$ range & $z_\mathrm{snp}$ & $\log_{10} M_0$ & $\bar{n}_g$ & $b_\mathrm{WiZ}$ \\
%%     \hline
%%     $\Delta z^\mathrm{Near}$ & 0.2 -- 0.6 & 0.44 & 12.17 & 7.44 & 1.01 \\
%%     $\Delta z^\mathrm{Mid}$  & 0.4 -- 0.8 & 0.60 & 12.28 & 5.86 & 1.14 \\
%%     $\Delta z^\mathrm{Far}$  & 0.6 -- 1.0 & 0.74 & 12.17 & 6.83 & 1.19 \\
%%     \hline
%%   \end{tabular}
%% \end{table*}

\section{Conclusion}
\begin{itemize}
\item We have presented the WiZ-COLA simulation, which consists of
  3600 simulations with $1296^3$ particles that covers the volume of $(600
  h^{-1} \mathrm{Mpc})^3$, and resolve haloes of mass $10^{12} h^{-1}
  \mathrm{Mpc}$, using our new parallelized COLA code. The simulation
  took only 200k core hours in total.

\item We generate 600 realisations of mock galaxy catalogues for the
  WiggleZ survey, and the BOSS CMASS galaxies in the overlap regions
  using HODs. We show that COLA can create mock HOD galaxies as
  accurate as GADGET $N$-body simulations for large-scale power
  spectra for wavelength $k \le 0.2 \,h\mathrm{Mpc}^{-1}$, both in
  real- and redshift-space.
  
\item The accuracy in peculiar velocity is the primary advantage of
  COLA simulations. We show that velocity power spectra are accurate
  within a per cent for $k \le 0.15 \,h \mathrm{Mpc}$ and 3 per cent
  for $0.2 \,h\mathrm{Mpc}$, and we expect a similar accuracy for the
  quadrupole moment of the galaxy power spectra in redshift space. The
  accuracy of the galaxy-velocity cross-power spectra and monopole
  moment of galaxy power spectra is better than 1 per cent.

\item Another benefit of the COLA approach is that the linear growth
  rate of matter fluctuation at large scales is determined much better
  than 1 per cent. This is not a great advantage for galaxy survey,
  because of the few-per-cent error in galaxy bias, but could be an
  advantage for gravitational lensing surveys.
\end{itemize}

%
% Acknowledgments, References
%
\section*{Acknowledgements}
This research was conducted by the Australian Research Council Centre
of Excellence for All-sky Astrophysics (CAASTRO), through project
number CE110001020. JK also acknowledges support of the European
Research Council through the Darklight ERC Advanced Research Grant
(\#291521). CB acknowledges the support of the Australian Research
Council through Future Fellowship awards FT110100639. Our numerical
computation was supported by the gSTAR national facility at
Swinburne University of Technology, the Flagship Allocation Scheme of
the NCI National Facility at the ANU, and the Texas Advanced Computing
Center (TACC) at The University of Texas at Austin. gSTAR is funded by
Swinburne and the Australian Governments Education Investment Fund.

\bibliographystyle{mn2e}
\bibliography{mock}
\appendix

%
% Appendix
%
\appendix
\section{Impact of initial condition}
\label{sec:appendix_initial_time}
We show the impact of our initial condition, which is given at
$a=0.5/n_\mathrm{step}$ for the velocity
(equation~\ref{eq:initial_time}), compared to the original one at
$a=0.1$ for both the position and velocity,
\begin{equation}
  \label{eq:initial_time_original}
  \bmath{x}^\mathrm{res}(t_1) = 0, \qquad \bmath{v}^\mathrm{res}(t_1) = 0.
\end{equation}
This original initial condition gives slightly better results,
although our initial condition is not problematic in theory. The
ansatz for COLA with $n_{LPT} = -2.5$ is tuned for the original
initial condition at $a=0.1$, and the same ansatz is probably not
optimal for our initial velocity at $a=0.05$.

\begin{figure}
  \centering
  \includegraphics[width=84mm]{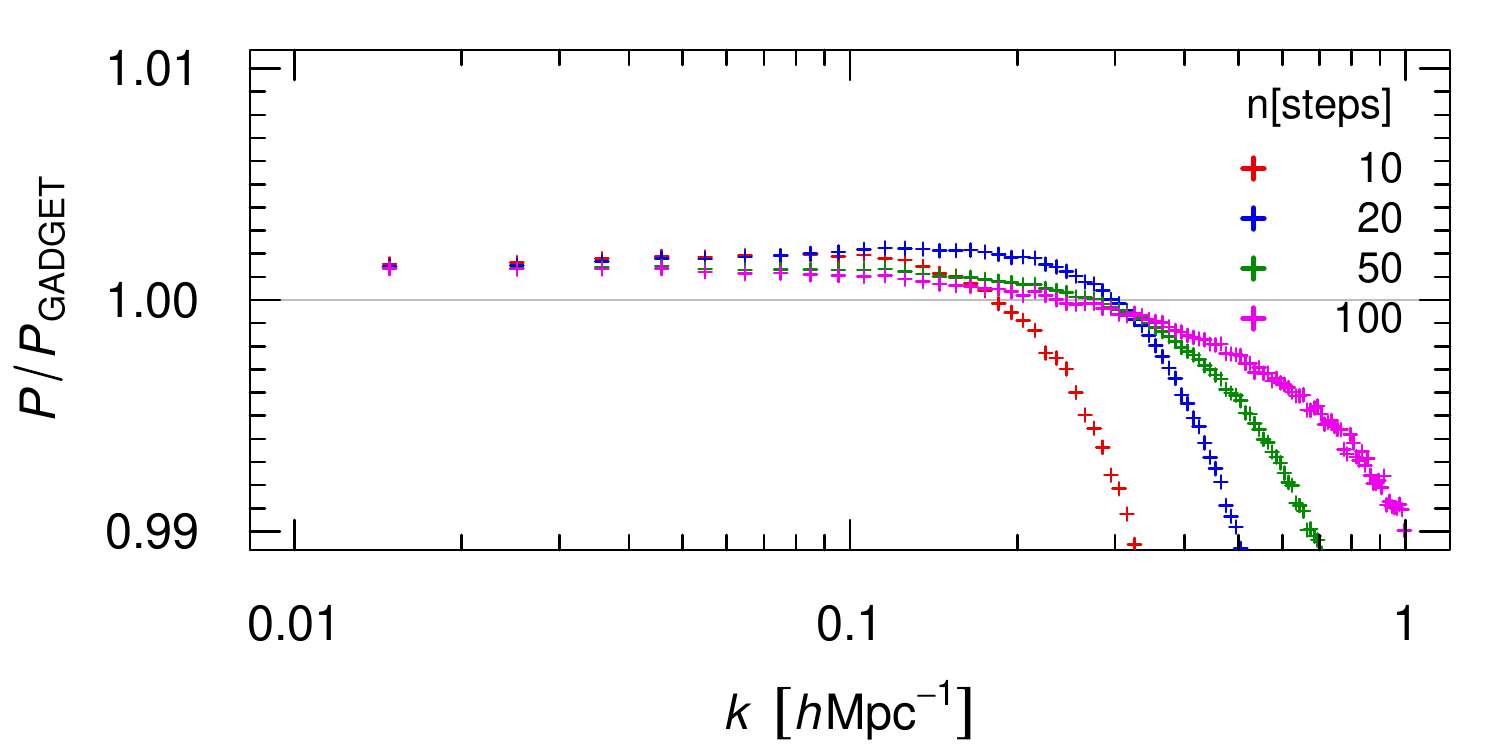}
  \caption{The matter power spectrum with the original initial
    condition, which gives slightly more accurate power spectrum than
    Fig.~\ref{fig:matter_power}.}
  \label{fig:power_spectra_appendix}
\end{figure}

In Fig.~\ref{fig:power_spectra_appendix}, we plot the ratio of the
matter power spectra to that of the GADGET $N$-body simulations at
$z=0.6$ for different number of steps with the original initial
condition. We divide the time equally in scale factor between 0 and 1,
$a(t_i)=i/n_\mathrm{step}$, for $n_\mathrm{step} = 10, 20, 50$, and
$100$.  The original initial condition gives better accuracy around $k
= 0.1\,h\mathrm{Mpc}^{-1}$, without the 2--3-per-cent excess in
Fig.~\ref{fig:matter_power}; the agreement is better than 1 per cent
for $k \le 0.3 \,h\mathrm{Mpc}^{-1}$. The range with accurate
  matter power expands as we increase the number of steps.

\begin{figure}
  \centering
    \includegraphics[width=84mm]{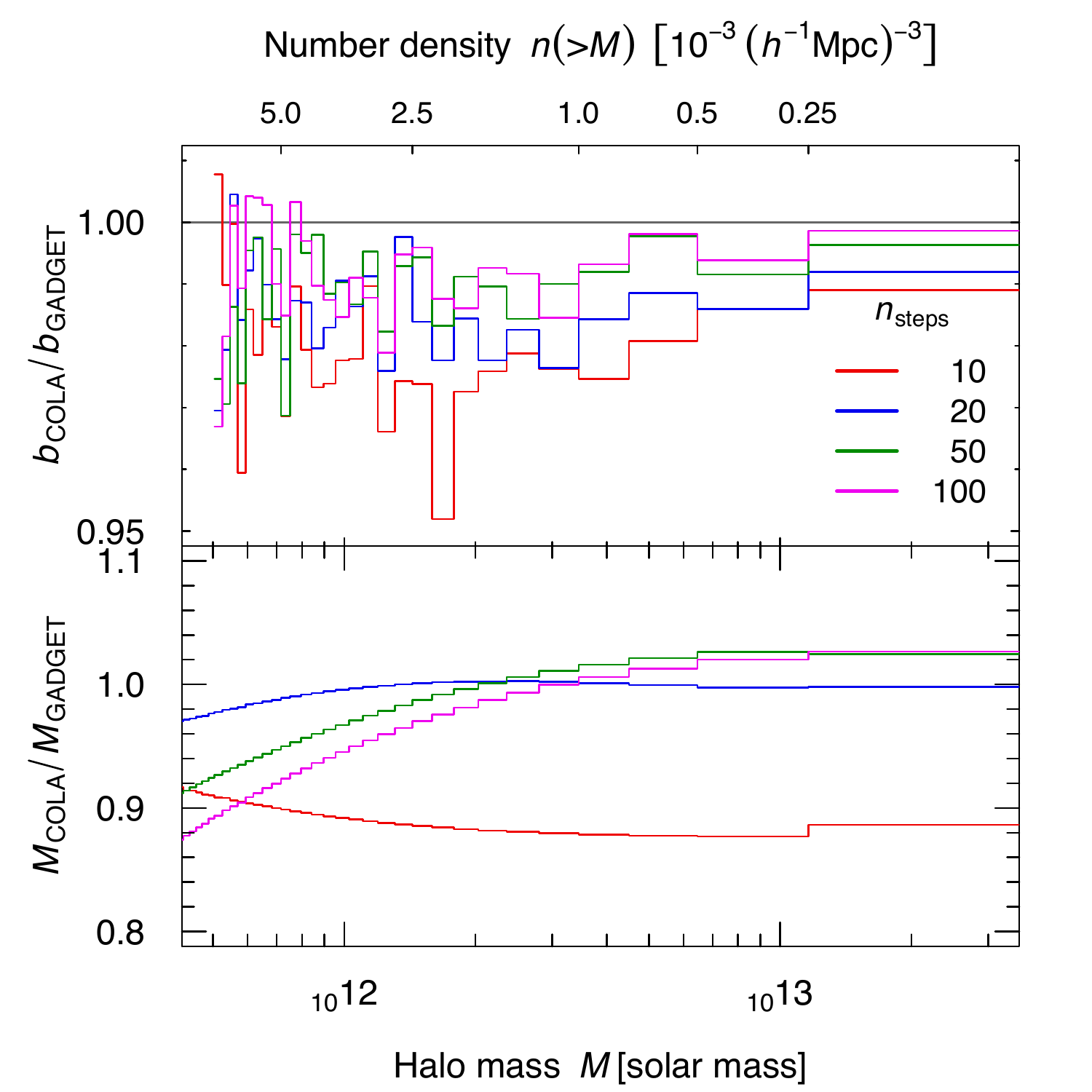}
  \caption{The precision of COLA halo bias (\textit{Upper panel}) and
    halo mass (\textit{Lower panel}) for various time steps. The
    original initial condition gives slightly better biases than
    Fig~\ref{fig:halo_bias}. The accuracy become about 1 per cent for
    100 steps, while the halo masses do not show monotonic
    convergence.}
  \label{fig:bias_appendix}
\end{figure}

In Fig.~\ref{fig:bias_appendix}, we plot the accuracy in the halo bias
and mass. The original initial condition gives a slight improvement
for the halo bias as well --- from 5-per-cent error in
Fig.~\ref{fig:halo_bias} to about 3 per cent for 10 time steps. We
split the haloes to groups with an equal number density of $10^{-4}
(h^{-1}\mathrm{Mpc})^{-3}$ by their mass and compute the halo bias, as
we did for Fig.~\ref{fig:halo_bias}.  The halo bias improves to about
1 per cent for 100 steps. The lower panel shows the mean halo mass in
each group. Our COLA simulations does not converge to the GADGET
simulation because we have the uniform PM grid for force computation,
and that causes an additional error in the halo formation independent
of time steps. The PM force recovers the correct force at a distance
of about 2.7 times the PM grid size, which corresponds to a virial
radius of a halo of mass $M_{200,m} = 5 \times 10^{12} h^{-1} M_\odot$
for our configuration; the limited force resolution below this scale
explains the deviation from the correct halo mass.

In this Appendix, we have shown that our excess in our
matter power spectrum was caused by our initial setup for the
velocity, and the accuracy of COLA simulations could improve slightly
by using the original initial condition.

\newpage

\begin{table*}
  \caption{We list the number of galaxies, $N_\mathrm{WiggleZ}$, the
    mean numbers of mock galaxies and their standard error in the mean
    for 3600 realisations, $\bar{N}_\mathrm{WiggleZ}$, and the survey
    volume in units of $10^7 (h^{-1}\mathrm{Mpc})^3$, for the six
    regions in the sky decomposed to 3 redshift bins. The cuboid is
    one of the box remappings listed in Table~\ref{table:remap}. The
    'overlap' is the fraction of the survey volume that overlaps in
    the periodic simulation box in per cent --- the overlapped volume
    consists of two copies of the same simulation volume. Since
    $\Delta^\mathrm{Mid}$ completely overlaps with the other two
    redshift bins, the total, in the final row, is the sum for Near
    and Far redshift bins. \label{table:hrs}}

  \begin{tabular}{llccrcc}
    \hline
    reg  & $\Delta z$ & $N_\mathrm{WiggleZ}$ & $\bar{N}_\mathrm{WiZ-COLA}$ &
    volume ($10^7$) & cuboid & overlap\\
    \hline
    1 hr  & Near & 6927  & $6927.63 \pm 3.3$ &  2.81 & $\sqrt{3}$ & 0\\
    1 hr  & Mid  & 9437  & $9436.5  \pm 3.4$ &  4.98 & $\sqrt{3}$ & 0\\
    1 hr  & Far  & 7880  & $7882.2  \pm 3.1$ &  7.12 & $\sqrt{3}$ & 0\\    
    3 hr  & Near & 8000  & $8000.3  \pm 3.6$ &  2.89 & $\sqrt{3}$ & 0\\
    3 hr  & Mid  & 10241 & $10240.7 \pm 3.6$ &  5.12 & $\sqrt{3}$ & 0\\
    3 hr  & Far  & 8756  & $8760.0  \pm 3.1$ &  7.33 & $\sqrt{3}$ & 0\\
    9 hr  & Near & 15128 & $15131.0 \pm 5.0$ &  4.82 & $\sqrt{3}$ & 0\\
    9 hr  & Mid  & 18978 & $18984.0 \pm 5.1$ &  8.53 & $\sqrt{3}$ & 0\\
    9 hr  & Far  & 11424 & $11418.6 \pm 3.4$ & 12.20 & $\sqrt{2}$ & 0.58\\
    11 hr & Near & 18019 & $18020.1 \pm 5.1$ &  6.25 & $\sqrt{3}$ & 0\\
    11 hr & Mid  & 22289 & $22299.2 \pm 4.8$ & 11.07 & $\sqrt{2}$ & 5.08\\
    11 hr & Far  & 13919 & $13894.9 \pm 3.3$ & 15.84 & $\sqrt{2}$ & 1.73\\
    15 hr & Near & 22309 & $22312.3 \pm 6.1$ &  7.12 & $\sqrt{3}$ & 0\\
    15 hr & Mid  & 30015 & $30024.6 \pm 6.1$ & 12.62 & $\sqrt{2}$ & 4.88\\
    15 hr & Far  & 19471 & $19428.3 \pm 4.4$ & 18.05 & $\sqrt{2}$ & 5.66\\
    22 hr & Near & 15884 & $15883.6 \pm 6.5$ &  3.55 & $\sqrt{3}$ & 0\\
    22 hr & Mid  & 16146 & $16142.7 \pm 5.4$ &  6.29 & $\sqrt{3}$ & 0\\
    22 hr & Far  & 11024 & $11025.9 \pm 3.8$ &  9.00 & $\sqrt{3}$ & 0\\
    \hline
    Total &      & 158741 &                  & 97.00 & & 1.7\\
    \hline
  \end{tabular}
\end{table*}

\bsp
\label{lastpage}
\end{document}